\documentclass[12pt,letterpaper]{article}

\usepackage{bbm}
\usepackage{amscd,amssymb,epsf}

\usepackage{textcomp}
\usepackage{hyperref}

\global\def\draftcontrol{0}

   \def\versionno{NS5 on Resolved Cone over  $Y^{p,q}$  }
\catcode`\@=11

\expandafter\ifx\csname draftcontrol\endcsname\relax\global\def\draftcontrol{0}
\fi

{\count255=\time\divide\count255 by 60
\xdef\hourmin{\number\count255}
\multiply\count255 by-60\advance\count255 by\time
\xdef\hourmin{\hourmin:\ifnum\count255<10 0\fi\the\count255}}
\def\draftdate{\number\month/\number\day/\number\year\ \ \ \hourmin }

\newcommand\makepapertitle{\par
  \begingroup
    \renewcommand\thefootnote{\@fnsymbol\c@footnote}%
    \def\@makefnmark{\rlap{\@textsuperscript{\normalfont\@thefnmark}}}%
    \long\def\@makefntext##1{\parindent 1em\noindent
            \hb@xt@1.8em{%
                \hss\@textsuperscript{\normalfont\@thefnmark}}##1}%
     \newpage
     \global\@topnum\z@   
     \@makepapertitle
     \thispagestyle{empty}\@thanks
  \endgroup
  \setcounter{footnote}{0}%
  \global\let\thanks\relax
  \global\let\makepapertitle\relax
  \global\let\@makepapertitle\relax
  \global\let\@thanks\@empty
  \global\let\@author\@empty
  \global\let\@date\@empty
  \global\let\@title\@empty
  \global\let\title\relax
  \global\let\author\relax
  \global\let\date\relax
  \global\let\and\relax
  \def\version{\let\version\@version\@gobble}
}
\def\@makepapertitle{%
  \newpage
   \ifnum\draftcontrol=1 {}
   \version\versionno
   \vskip 3em%
   \else
   \hfill\hbox to 3cm {\parbox{4cm}{\@pubnum}\hss}%
   \vskip 3em%
   \fi
   \begin{center}%
   \let \footnote \thanks
     {\LARGE {\@title}}%
     \vskip 1.5em%
     {\normalsize
       \lineskip .5em%
       \begin{tabular}[t]{c}%
         \@author
       \end{tabular}\par}%
     \vskip 1.5em%
     {\@bstract}%
     \end{center}%
     \vskip 1.5em
     \@date%
   \par
}

\gdef\@pubnum{}
\def\pubnum#1{%
  \gdef\@pubnum{#1}}

\gdef\@bstract{}
\def\Abstract#1{%
  \gdef\@bstract{%
   \parbox{\textwidth-0pc}{%
   \centerline{\bf Abstract}\penalty1000%
\noindent
\renewcommand\baselinestretch{1.0}%
{#1}}}
}

\def\ps@paper{\let\@mkboth\@gobbletwo%
     \ifnum\draftcontrol=1
        \def\@oddfoot{\hbox to \textwidth{\tiny \versionno \hfil\tiny\draftdate}%
        \hskip -\textwidth \hbox to \textwidth{\hfil\rm\thepage\hfil}}%
     \else\def\@oddfoot{\hbox to \textwidth{\hfil\rm\thepage\hfil}}
     \fi
     \let\@evenfoot\@oddfoot
}

\def\@version#1{\ifnum\draftcontrol=1
\typeout{}\typeout{#1}\typeout{}
\vskip3mm\centerline{\hbox{\fbox{\normalsize{\tt DRAFT -- #1 -- }
                   {\draftdate}}}}\vskip3mm
\fi}
\let\version\@version
\long\def\eqlabel#1{\ifnum\draftcontrol=1
                    \tag@false  
                    \tag*{(\theequation) \hbox to -0.2cm{\hspace{0cm}\small{#1}\hss}}
                    \refstepcounter{equation}
                    \edef\@currentlabel{\theequation}
                    \ltx@label{#1}          
                    \else
                    \label{#1}
                    \fi
                    }
\let\st@bibitem\@bibitem
\let\st@lbibitem\@lbibitem
\ifnum\draftcontrol=1
  \def\@bibitem#1{%
    \st@bibitem{#1}\a@@label{#1}\ignorespaces}
  \def\@lbibitem[#1]#2{%
    \st@lbibitem[#1]{#2}\a@@label{#2}\ignorespaces}
  \def\a@@label#1{%
    \gdef\a@lab{\smash{\normalfont\small#1}}
    \ifvmode
      \if@inlabel
        \global\setbox\@labels\hbox{%
          \llap{\a@lab\let\a@lab\relax
                \kern\@totalleftmargin\kern\marginparsep}%
          \box\@labels}%
      \fi
    \fi}
\fi

\usepackage{amsmath,amssymb,array,calc,rotating,epsfig,psfrag}
\usepackage[nosort]{cite}

\ifnum\draftcontrol=1
\fi

\renewcommand\baselinestretch{1.25}
\renewcommand\section{\@startsection {section}{1}{\z@}%
                                   {-3.5ex \@plus -1ex \@minus -.2ex}%
                                   {2.3ex \@plus.2ex}%
                                   {\normalfont\large\bfseries}}
\renewcommand\subsection{\@startsection{subsection}{2}{\z@}%
                                   {-3.25ex\@plus -1ex \@minus -.2ex}%
                                   {1.5ex \@plus .2ex}%
                                   {\normalfont\normalsize\bfseries}}
\renewcommand\subsubsection{\@startsection{subsubsection}{3}{\z@}%
                                   {-3.25ex\@plus -1ex \@minus -.2ex}%
                                   {1.5ex \@plus .2ex}%
                                   {\normalfont\normalsize\it}}
\renewcommand\paragraph{\@startsection{paragraph}{4}{\z@}%
                                   {-3.25ex\@plus -1ex \@minus -.2ex}%
                                   {1.5ex \@plus .2ex}%
                                   {\normalfont\normalsize\bf}}

\def\revise#1       {\raisebox{-0em}{\rule{3pt}{1em}}%
                     \marginpar{\raisebox{.5em}{\vrule width3pt\
                     \vrule width0pt height 0pt depth0.5em
                     \hbox to 0cm{\hspace{0cm}{%
                     \parbox[t]{4em}{\raggedright\footnotesize{#1}}}\hss}}}}

\def\cala         {{\cal A}}
\def\calA         {{\mathfrak A}}
\def\calAbar      {{\underline \calA}}
\def\calb         {{\cal B}}
\def\calc         {{\cal C}}
\def\cald         {{\cal D}}
\def\cale         {{\cal E}}
\def\calf         {{\cal F}}
\def\calg         {{\cal G}}
\def\calG         {{\mathfrak G}}
\def\calh         {{\cal H}}
\def\cali         {{\cal I}}
\def\calj         {{\cal J}}
\def\calk         {{\cal K}}
\def\call         {{\cal L}}
\def\calm         {{\cal M}}
\def\caln         {{\cal N}}
\def\calo         {{\cal O}}
\def\calp         {{\cal P}}
\def\calq         {{\cal Q}}
\def\calr         {{\cal R}}
\def\cals         {{\cal S}}
\def\calt         {{\cal T}}
\def\calu         {{\cal U}}
\def\calv         {{\cal V}}
\def\calw         {{\cal W}}

\def\complex      {{\mathbb C}}
\def\naturals     {{\mathbb N}}
\def\projective   {{\mathbb P}}
\def\rationals    {{\mathbb Q}}
\def\reals        {{\mathbb R}}
\def\zet          {{\mathbb Z}}

\def\del          {\partial}

\def\ee           {{\rm e}}

\def\tr           {\mathop{\rm Tr}}

\def\de#1#2{{\rm d}^{#1}\!#2\,}

\def\sqr#1#2{{\vcenter{\vbox{\hrule height.#2pt
 \hbox{\vrule width.#2pt height#1pt \kern#1pt
 \vrule width.#2pt}\hrule height.#2pt}}}}

\def\a{\alpha}
\def\b{\beta}
\def\r{\rho}

\def\la{\lambda}

\def\be{\begin{equation}}
\def\ee{\end{equation}}

\def\m{\mu}
\def\g{\gamma}
\def\l{\lambda}
\def\n{\nu}

\catcode`\@=12

\begin{document}



\def \el {{\ell}}
\def \KK {{\cal  K}}
\def \K {{\rm K}}
\def \tz{\tilde{z}}

\def \ci {\cite}
\newcommand{\rf}[1]{(\ref{#1})}
\def \la {\label}
\def \const {{\rm const}}

\def \ov {\over}
\def \ha {\textstyle { 1\ov 2}}
\def \we { \wedge}
\def \P { \Phi}
\def\ep {\epsilon}
\def \ab {{A^2 \ov B^2}}
\def \ba {{B^2 \ov A^2}}
\def \tv   {{1 \ov 12}}
\def \go { g_1}
\def \gd { g_2}
\def \gt { g_3}
\def \gc { g_4}
\def \gp { g_5}
\def \F {{\cal F}}
\def \del { \partial}
\def \t {\theta}
\def \p {\phi}
\def \te {\tilde \epsilon}
\def \ps {\psi}
\def \x {{x_{11}}}

\def\br{\bar{\rho}}
\newcounter{subequation}[equation]

\def\pa{\partial}
\def\e{\epsilon}
\def\rt{\rightarrow}
\def\tr{{\tilde\rho}}
\newcommand{\beq}{\begin{equation}}
\newcommand{\eeq}{\end{equation}}
\newcommand{\eel}[1]{\label{#1}\end{equation}}
\newcommand{\bea}{\begin{eqnarray}}
\newcommand{\eea}{\end{eqnarray}}
\newcommand{\bear}{\begin{eqnarray}}
\newcommand{\eear}{\end{eqnarray}}
\newcommand{\eeal}[1]{\label{#1}\end{eqnarray}}
\newcommand{\LL}{e^{2\lambda(r)}}
\newcommand{\NN}{e^{2\nu(r)}}
\newcommand{\PP}{e^{-2\phi(r)}}
\newcommand{\non}{\nonumber \\}
\newcommand{\CR}{\non\cr}
\newcommand{\rc}{\nonumber\\}

\makeatletter

\def\thesubequation{\theequation\@alph\c@subequation}
\def\@subeqnnum{{\rm (\thesubequation)}}
\def\slabel#1{\@bsphack\if@filesw {\let\thepage\relax
   \xdef\@gtempa{\write\@auxout{\string
      \newlabel{#1}{{\thesubequation}{\thepage}}}}}\@gtempa
   \if@nobreak \ifvmode\nobreak\fi\fi\fi\@esphack}
\def\subeqnarray{\stepcounter{equation}
\let\@currentlabel=\theequation\global\c@subequation\@ne
\global\@eqnswtrue \global\@eqcnt\z@\tabskip\@centering\let\\=\@subeqncr

$$\halign to \displaywidth\bgroup\@eqnsel\hskip\@centering
  $\displaystyle\tabskip\z@{##}$&\global\@eqcnt\@ne
  \hskip 2\arraycolsep \hfil${##}$\hfil
  &\global\@eqcnt\tw@ \hskip 2\arraycolsep
  $\displaystyle\tabskip\z@{##}$\hfil
   \tabskip\@centering&\llap{##}\tabskip\z@\cr}
\def\endsubeqnarray{\@@subeqncr\egroup
                     $$\global\@ignoretrue}
\def\@subeqncr{{\ifnum0=`}\fi\@ifstar{\global\@eqpen\@M
    \@ysubeqncr}{\global\@eqpen\interdisplaylinepenalty \@ysubeqncr}}
\def\@ysubeqncr{\@ifnextchar [{\@xsubeqncr}{\@xsubeqncr[\z@]}}
\def\@xsubeqncr[#1]{\ifnum0=`{\fi}\@@subeqncr
   \noalign{\penalty\@eqpen\vskip\jot\vskip #1\relax}}
\def\@@subeqncr{\let\@tempa\relax
    \ifcase\@eqcnt \def\@tempa{& & &}\or \def\@tempa{& &}
      \else \def\@tempa{&}\fi
     \@tempa \if@eqnsw\@subeqnnum\refstepcounter{subequation}\fi
     \global\@eqnswtrue\global\@eqcnt\z@\cr}
\let\@ssubeqncr=\@subeqncr
\@namedef{subeqnarray*}{\def\@subeqncr{\nonumber\@ssubeqncr}\subeqnarray}

\@namedef{endsubeqnarray*}{\global\advance\c@equation\m@ne
                           \nonumber\endsubeqnarray}

\makeatletter \@addtoreset{equation}{section} \makeatother
\renewcommand{\theequation}{\thesection.\arabic{equation}}

\def \ci {\cite}
\def \la {\label}
\def \const {{\rm const}}
\catcode`\@=11

\newcount\hour
\newcount\minute
\newtoks\amorpm \hour=\time\divide\hour by 60\minute
=\time{\multiply\hour by 60 \global\advance\minute by-\hour}
\edef\standardtime{{\ifnum\hour<12 \global\amorpm={am}
        \else\global\amorpm={pm}\advance\hour by-12 \fi
        \ifnum\hour=0 \hour=12 \fi
        \number\hour:\ifnum\minute<10
        0\fi\number\minute\the\amorpm}}
\edef\militarytime{\number\hour:\ifnum\minute<10 0\fi\number\minute}

\def\draftlabel#1{{\@bsphack\if@filesw {\let\thepage\relax
   \xdef\@gtempa{\write\@auxout{\string
      \newlabel{#1}{{\@currentlabel}{\thepage}}}}}\@gtempa
   \if@nobreak \ifvmode\nobreak\fi\fi\fi\@esphack}
        \gdef\@eqnlabel{#1}}
\def\@eqnlabel{}
\def\@vacuum{}
\def\marginnote#1{}
\def\draftmarginnote#1{\marginpar{\raggedright\scriptsize\tt#1}}
\overfullrule=0pt

 \def \lc {light-cone\ }

\def\draft{
        \pagestyle{plain}
        \overfullrule=2pt
        \oddsidemargin -.5truein
        \def\@oddhead{\sl \phantom{\today\quad\militarytime} \hfil
        \smash{\Large\sl DRAFT} \hfil \today\quad\militarytime}
        \let\@evenhead\@oddhead
        \let\label=\draftlabel
        \let\marginnote=\draftmarginnote
        \def\ps@empty{\let\@mkboth\@gobbletwo
        \def\@oddfoot{\hfil \smash{\Large\sl DRAFT} \hfil}
        \let\@evenfoot\@oddhead}

\def\@eqnnum{(\theequation)\rlap{\kern\marginparsep\tt\@eqnlabel}
        \global\let\@eqnlabel\@vacuum}  }

\renewcommand{\rf}[1]{(\ref{#1})}
\renewcommand{\theequation}{\thesection.\arabic{equation}}
\renewcommand{\thefootnote}{\fnsymbol{footnote}}

\newcommand{\newsection}{    
\setcounter{equation}{0}
\section}

\textheight = 22truecm
\textwidth = 17truecm
\hoffset = -1.3truecm
\voffset =-.9truecm

\def \tx {\textstyle}
\def \tix{\tilde{x}}
\def \bi{\bibitem}

\def \ov {\over}
\def \ha {\textstyle { 1\ov 2}}
\def \we { \wedge}
\def \P { \Phi} \def\ep {\epsilon}
\def \ab {{A^2 \ov B^2}}
\def \ba {{B^2 \ov A^2}}
\def \tv   {{1 \ov 12}}
\def \go { g_1}\def \gd { g_2}\def \gt { g_3}
\def \gc { g_4}\def \gp {
g_5}
\def \F {{\cal F}}
\def \del { \partial}
\def \t {\theta}
\def \p {\phi}
\def \ep {\epsilon}
\def \ps {\psi}

\def \LL{{\cal L}}
\def\o{\omega}
\def\O{\Omega}
\def\e{\epsilon}
\def\pd{\partial}
\def\pdz{\partial_{\bar{z}}}
\def\bz{\bar{z}}
\def\e{\epsilon}
\def\m{\mu}
\def\n{\nu}
\def\a{\alpha}
\def\b{\beta}
\def\g{\gamma}
\def\G{\Gamma}
\def\d{\delta}
\def\r{\rho}
\def\bx{\bar{x}}
\def\by{\bar{y}}
\def\bm{\bar{m}}
\def\bn{\bar{n}}
\def\s{\sigma}
\def\na{\nabla}
\def\D{\Delta}
\def\l{\lambda}
\def\te{\theta} \def \t {\theta}
\def\ta {\tau}
\def\na{\bigtriangledown}
\def\p{\phi}
\def\L{\Lambda}
\def\hR{\hat R}
\def\ch{{\cal H}}
\def\ep{\epsilon}
\def\bj{\bar{J}}
\def \foot{ \footnote}
\def\be{\begin{equation}}
\def\ee{\end{equation}}
\def \P {\Phi}
\def\un{\underline{n}}
\def\ur{\underline{r}}
\def\um{\underline{m}}
\def \ci {\cite}
\def \g {\gamma}
\def \G {\Gamma}
\def \k {\kappa}
\def \l {\lambda}
\def \L {{L}}
\def \Tr {{\rm Tr}}
\def\apr{{A'}}
\def \m {\mu}
\def \n {\nu}
\def \W{{\cal W}}
\def \eps {\epsilon}
\def \ha{{
 { 1 \ov 2}} }
\def \de{{
{ 1 \ov 9}} }
\def \si{{
 { 1 \ov 6}} }
\def \fo{{
{ 1 \ov 4}} }
\def \ei{{
{ 1 \ov 8}} }
\def \rt {{\tx { \ta \ov 2}}}
\def \rr {{\bar \rho}}
\def\cF{{\cal{F}}}

\def\D{\Delta}
\def\l{\lambda}
\def\L{\Lambda}
\def\te{\theta}
\def\g{\gamma}
\def\Te{\Theta}
\def\tw{\tilde{w}}

\def\sn{\rm sn}
\def\cn{\rm cn}
\def\dn{\rm dn}
\def\a{\alpha}
\def\as{\asymp}
\def\ap{\approx}
\def\b{\beta}
\def\da{\dagger}
\def\d{\delta}
\def\D{\Delta}
\def\ep{\epsilon}
\def\eq{\equiv}
\def\f{\frac}
\def\g{\gamma}
\def\G{\Gamma}
\def\hs{\hspace}
\def\hX{\hat{X}}
\def\hY{\hat{Y}}
\def\k{\kappa}
\def\lf{\left}
\def\l{\lambda}
\def\L{\Lambda}
\def\na{\nabla}
\def\nn{\nonumber}
\def\o{\omega}
\def\O{\Omega}
\def\p{\phi}
\def\pa{\partial}
\def\P{\Phi}
\def\r{\rho}
\def\ra{\rightarrow}
\def\ri{\right}
\def\s{\sigma}
\def\S{\Sigma}
\def\si{\simeq}
\def\st{\star}
\def\t{\theta}
\def\ti{\tilde}
\def\tm{\times}
\def\tr{\textrm}
\def\T{\Theta}
\def\u{\upsilon}
\def\U{\Upsilon}
\def\v{\varepsilon}
\def\vh{\varphi}
\def\vp{\varpi}
\def\vr{\varrho}
\def\vs{\varsigma}
\def\vt{\vartheta}
\def\w{\wedge}
\def\z{\zeta}
\def\bep{\bar{\epsilon}}
\def\hw{\hat{\omega}}
\def\cala         {{\cal A}}
\def\calA         {{\mathfrak A}}
\def\calAbar      {{\underline \calA}}
\def\calb         {{\cal B}}
\def\calc         {{\cal C}}
\def\cald         {{\cal D}}
\def\cale         {{\cal E}}
\def\calf         {{\cal F}}
\def\calg         {{\cal G}}
\def\calG         {{\mathfrak G}}
\def\calh         {{\cal H}}
\def\cali         {{\cal I}}
\def\calj         {{\cal J}}
\def\calk         {{\cal K}}
\def\call         {{\cal L}}
\def\calm         {{\cal M}}
\def\caln         {{\cal N}}
\def\calo         {{\cal O}}
\def\calp         {{\cal P}}
\def\calq         {{\cal Q}}
\def\calr         {{\cal R}}
\def\cals         {{\cal S}}
\def\calt         {{\cal T}}
\def\calu         {{\cal U}}
\def\calv         {{\cal V}}
\def\calw         {{\cal W}}
\def\calz         {{\cal Z}}
\def\complex      {{\mathbb C}}
\def\naturals     {{\mathbb N}}
\def\projective   {{\mathbb P}}
\def\rationals    {{\mathbb Q}}
\def\reals        {{\mathbb R}}
\def\zet          {{\mathbb Z}}

\def\hzero{\hat{0}}
\def\ha{\hat{a}}
\def\hb{\hat{b}}
\def\hc{\hat{c}}
\def\hd{\hat{d}}
\def\he{\hat{e}}

\def\hone{\hat{1}}
\def\htwo{\hat{2}}
\def\hthree{\hat{3}}
\def\hz{\hat{z}}
\def\hteone{\hat{\theta}_1}
\def\htetwo{\hat{\theta}_2}
\def\hpone{\hat{\phi}_1}
\def\hptwo{\hat{\phi}_2}
\def\hpsi{\hat{\psi}}


\topmargin=0.50in

\date{}

\begin{titlepage}

\version\versionno

\begin{flushright}
 MCTP-10-21\\
 UTTG-05-10\\
 TIFR/TH/10-19
\end{flushright}

\begin{center}
%
 {\Large \bf  Toward NS5 Branes on the
Resolved Cone over $Y^{p,q}$}
\vskip .2cm

\vskip .7 cm
{\bf  Elena C\'aceres${}^1$, \,\,\,\, Manavendra N. Mahato,${}^2$}\\
\vspace{.2cm}

{\bf Leopoldo A. Pando Zayas${}^3$ and Vincent G. J. Rodgers${}^4$}

\vskip .2 cm

\end{center}

\begin{center}
\begin{tabular}{cc}
{\it \small ${}^1$Facultad de Ciencias } & {\it \small ${}^2$ Department of Theoretical
Physics}\\
{ \it \small Universidad de  Colima}& { \it \small Tata Institute of Fundamental Research}\\
{\it \small Bernal D\'\i az del Castillo 340}& {\it\small  Mumbai 400 005, India}\\
{\it \small Col. Villas San Sebasti\'an } &\\
{\it \small Colima, Colima 28045, M\'exico }& \\
&\\
{\it \small ${}^3$Michigan Center for Theoretical
Physics}& {\it \small ${}^4$Department of Physics and Astronomy}\\
{ \it \small Randall Laboratory of Physics} & { \it \small The University of Iowa}\\
{\it \small The University of Michigan} &{\it \small Iowa City, IA 52242, USA }\\
{\it \small Ann Arbor, MI 48109-1040, USA }& \\
\end{tabular}
\end{center}


\begin{abstract}
Motivated by recent developments in the understanding of the connection between five branes on resolved geometries and the corresponding generalizations
of complex deformations in the context of the warped resolved deformed conifold, we consider the construction of five branes
solutions on the resolved cone over $Y^{p,q}$ spaces. We establish the existence of supersymmetric five branes solutions wrapped on two-cycles of the resolved cone over $Y^{p,q}$ in the probe limit. We then use
calibration techniques to begin the  construction of fully back-reacted five branes; we present an ansatz and the corresponding
equations of motion. Our results establish  a detailed framework to
study  back-reacted five branes wrapped on the resolved cone  over $Y^{p,q}$ and as a first step we find explicit solutions and construct
an asymptotic expansion with the expected properties.
\end{abstract}

\end{titlepage}
\setcounter{page}{1} \renewcommand{\thefootnote}{\arabic{footnote}}
\setcounter{footnote}{0}

\def \N{{\cal N}}
\def \ov {\over}

\tableofcontents

\section{Introduction}

The AdS/CFT correspondence provides a powerful tool to attack very important questions of strong coupling dynamics using gravitational duals. Particularly interesting is the class of supergravity backgrounds dual to confining theories containing ${\cal N}=1$ supersymmetric Yang-Mills (SYM). The original prototypes of these solutions are the Klebanov-Strassler solution (KS) \cite{Klebanov:2000hb} based on the deformed conifold and the
Maldacena-N\'u\~nez solution (MN) \cite{Maldacena:2000yy} based on an NS5 brane wrapping a two-cycle. Significant  progress has taken place in the  past ten years since
those seminal works appeared. One very important step in the construction of supergravity solutions was the first attempt
to relate the Maldacena-N\'u\~nez  and the Klebanov-Strassler solutions by means of an interpolating Ansatz presented in
Papadopoulos-Tseytlin \cite{Papadopoulos:2000gj}; it was shown that  both
solutions can be extracted from a single one-dimensional action. This idea was taken a step further in \cite{Butti:2004pk} where $SU(3)$-structure techniques were used
to construct the one-parameter family that realizes the interpolation.
 In a recent paper Maldacena and Martelli \cite{Maldacena:2009mw} have
further interpreted the results in \cite{Butti:2004pk} using a chain of dualities  and found a more complete picture that includes a supergravity
realization of geometric transition between the deformed conifold with fluxes and the resolved conifold with branes.
Another avenue of progress was started by Casero, N\'u\~nez and Paredes in \cite{Casero:2006pt} where they tackled  the problem of adding dynamical flavor to the Chamsedine-Volkov-Maldacena-N\'u\~nez (CVMN) background
\cite{Chamseddine:1997nm,Chamseddine1998,Maldacena:2000yy}. This line of research was further developed in \cite{Casero2007,Caceres2007,HoyosBadajoz:2008fw,Bigazzi2008, Bigazzi2008a,Caceres2009,Bigazzi2009}.
Finally, in \cite{Gaillard:2010qg}, exploiting an interpolation discussed in \cite{Minasian:2009rn},  the authors discuss a  solution generating technique that   can be used to generalize the deformed resolved conifold solution of \cite{Butti:2004pk}.

Despite all these  advancements, no {\it new} family of supergravity solutions containing a sector dual to ${\cal N}=1$ SYM has been constructed. One hopeful venue was
introduced with the construction of $Y^{p,q}$ spaces \cite{Gauntlett:2004yd,Gauntlett:2004hh}. The study of field theory duals
to $AdS_5\times Y^{p,q}$ spaces has produced interesting generalizations of the conifold theories. The dual field theory is rich and its understanding helped
clarified key aspects of the correspondence. The field theory dual to $AdS_5\times Y^{p,q}$ spaces was worked out in \cite{Bertolini:2004xf} and \cite{Benvenuti:2004dy}. Further field theoretic analysis of the corresponding cascading quivers indicates that  supersymmetry
is broken \cite{Bertolini:2005di}, \cite{Berenstein:2005xa}  \cite{Franco:2005zu}. This result
fits nicely with the fact that Calabi-Yau
deformations of the cone over $Y^{p,q}$ are obstructed \cite{Altmann:9403004,Altmann:9405008} and is one of the reasons why the  study of these models was not pursued further.
However, in view of
recent work \cite{Butti:2004pk,Maldacena:2009mw,Casero:2006pt,Gaillard:2010qg},  a logical alternative is to attack the problem from the point of
view of wrapping fivebranes which avoids altogether the need for a Calabi-Yau structure and relies only on the more general concept of  $SU(3)$ structure. This is what we attempt to initiate in this manuscript.

From the gravity point of view, the fact that there is no complex deformation of the cone over $Y^{p,q}$ \cite{Altmann:9403004,Altmann:9405008} means that there is no direct analog of the KS solution, that is, there is no solution of $D3$ and $D5$ built around a conformal Calabi-Yau that has a noncollapsing $S^3$ at the tip despite the perturbative evidence gathered in \cite{Herzog:2004tr} and more importantly in \cite{Burrington:2005zd}. Recent work by Maldacena and Martelli indicates  that the noncollapsing $S^3$ could appear
also as a consequence of the backreaction of the fivebranes. The non-K\"ahler analog of the deformed cone over $Y^{p,q}$ could thus be a solution with $H_3$ which preserves ${\cal N}=1$ supersymmetry.
Could the addition of branes or fluxes smoothly connect the resolved $Y^{p,q}$ and the ``appropriate'' notion of deformation? This would be the generalization of the situation in the conifold that was argued by Vafa in \cite{Vafa:2000wi} and realized purely in the supergravity context by  Maldacena-Martelli
\cite{Maldacena:2009mw}. The hope is to search starting the class of $SU(3)$ structure solutions rather than in the class of $SU(3)$ holonomy.

In the present work we aim to construct a  supergravity solution corresponding to backreacting NS5 branes wrapping
a two-cycle in a resolution of the cone over  $Y^{p,q}$. To gather evidence for the existence of such a solution we first find (section \ref{sec:probe})  a probe brane solution corresponding to a D5 brane on the resolved cone over $Y^{p,q}$. The existence of such D5 brane probe suggests the existence of a full back-reacted supergravity solution for D5 which we can, in turn, S-dualize to obtain the NS5 solution we seek. With this evidence in hand we proceed in section \ref{sec:NS5sol}
to  obtain the  equations of motion that  define the background. We show that these partial differential equations are consistent, study the asymptotic behavior and examine one particular case.
We consider the present work a first step in the study of branes on the resolved cone over $Y^{p,q}$ ; there are a myriad of issues to explore
and we comment on some of them in the conclusions.

\section{Review of $Y^{p,q}$ metric and the resolved cone over $Y^{p,q}$}\label{Sec:Review}
The starting point of our analysis are the  $Y^{p,q}$ spaces  whose metric was presented in  \cite{Gauntlett:2004yd}:
\bea
\label{cone-original}
ds^2 &=&\frac{1-cy}{6}\left(d\theta^2 +\sin^2\theta d\phi^2\right)+\frac{1}{w(y)q(y)}dy^2 + \frac{q(y)}{9} (d\psi- \cos\theta d\phi)^2   \\
&+& w(y)\left(d\a +\frac{ac -2y +y^2 c}{6(a-y^2)}(d\psi -\cos\theta d\phi)\right)^2,
\eea
with
\be
\quad w(y)=\frac{2(a-y^2)}{1-cy}, \quad q(y) = \frac{a-3y^2 + 2c y^3}{a-y^2}.
\ee
This is a two-parameter $(a,c)$  family of metrics. Typically if $c\ne 0$ it can be set to $c=1$  by rescaling  $y$.

This family of metrics contains $S^5$ and $T^{1,1}$ as particular limits. For us, it will be particularly interesting to consider the $T^{1,1}$ limit which has been explained in section 5 of \cite{Gauntlett:2004yd}. In this limit one requires $c\to 0$ in the standard notation of \cite{Gauntlett:2004yd}, we also need $a=3,\,\, y=\cos\omega$ and $\alpha =\nu/6$. The $Y^{p,q}$ metric  then becomes
\be
ds^2|_{c\to 0}=\frac16\left(d\theta^2 +\sin^2\theta d\phi^2 +d\omega^2 +\sin^2\omega d\nu^2 \right) + \frac19 \left(d\psi - \cos\theta d\phi - \cos\omega d\nu\right)^2,
\ee
which is readily recognized as the metric on $T^{1,1}$ as described in \cite{Candelas:1989js}.

\subsection{The resolved cone over  $Y^{p,q}$}
The $Y^{p,q}$ metrics are Sasaki-Einstein and therefore a cone over them is Calabi-Yau. A natural question is whether this Calabi-Yau space admits resolutions. The answer to that question is in the positive as opposed to the answer about complex deformation which is answered in
the negative \cite{Altmann:9403004,Altmann:9405008}. Following  the notation of \cite{Candelas:1989js} we will denote the resolved
cone over $Y^{p,q}$ as $\check{C}(Y^{p,q})$. The metric on the resolved cone over $Y^{p,q}$ was obtained explicitly in \cite{Oota:2006pm,Lu:2006cw} and further elaborations and extensions considering weighted projective $\mathbb{CP}^1$ were presented in \cite{Martelli:2007pv}. The metric in question is

\begin{eqnarray}
\label{CYM}
ds^2&=&\frac{(1-x)(1-y)}{4}(d\theta^2+\sin^2 \theta d\phi^2)
+\frac{(y-x)}{4X(x)}dx^2+\frac{(y-x)}{4Y(y)}dy^2 \\
&+&\frac{X(x)}{(y-x)}(d\tau +
(1-y)(d\psi-\cos \theta d\phi))^2\nonumber \\
&+&\frac{Y(y)}{(y-x)}(d\tau +
(1-x)(d\psi -\cos \theta d\phi))^2,\nonumber
\end{eqnarray}
where
\be
X(x)=x-1+\f{2}{3}(x-1)^2+\f{2\mu}{x-1},~~Y(y)=1-y-\f{2}{3}(1-y)^2-\f{2\nu}{1-y}
\ee
with two parameters $\mu$ and $\nu$.

As explained in  \cite{Martelli:2007pv}, to extend equation (\ref{CYM})
to a globally well defined non-compact manifold  we have to take
$y_1< y< y_2 $
where $y_1$ and $y_2$ are two consecutive roots of $Y(y)$.
Requiring  $0 \leq \nu \leq 1/6$ guarantees that  $y_1< y_2
< 1$ and
 $y_1 \leq 0$ while $y_2 \geq 0$. Thus, $Y(y)>0$,\quad
$\forall  y \in ( y_1, y_2)$.
We take $x$ to be non-compact and denote two consecutive
roots of $X(x)$ by $x_+$ and $x_-$.
It was shown in \cite{Martelli:2007pv} that $X(x)>0$,\quad
$\forall  x \in (-\infty, x_-)  \cup  (x_+ , \infty)$. As is
clear from (\ref{CYM}),
 we focus on the case where the resolution is obtained by
blowing up a $\mathbb{CP}^1$, referred to as ``small partial
resolutions  I'' in   \cite{Martelli:2007pv}. For this type of resolution we have $x_-=y_1$ which requires
$\mu= -\nu$. Thus, throughout this work we will consider
\be
-\infty < x < y_1 < 0,  \quad \quad   y_1 < y <y_2, \quad
\quad \mu= -\nu. \label{eq2.7}
\ee
We focus on the $\mathbb{CP}^1$ case although we presume that much of what we say can be adapted to the projective $\mathbb{CP}^1$ resolution presented in \cite{Martelli:2007pv}.

The above metric can be written using the following sechsbein $ds^2 =\delta_{ab} e^a e^b$:
\bea
\label{eq:Sechsbein}
e^1&=&\f{\sqrt{(1-x)(1-y)}}{2}(\cos(2(\tau + \psi))d\theta - \sin(2(\tau+\psi)) \sin\t d\phi),\\
e^2&=&\f{\sqrt{(1-x)(1-y)}}{2}(\sin(2(\tau+\psi)) d\theta +\cos(2(\tau+\psi)) \sin\theta d\phi),\nn \\
e^3&=&\sqrt{\f{X(x)}{(y-x)}}(d\tau+(1-y)(d\psi + A)), \quad e^4=-\sqrt{\f{y-x}{4Y(y)}}dy\nn \\
e^5&=&\sqrt{\f{Y(y)}{(y-x)}}(d\tau +(1-x)(d\psi +A)),  \quad e^6=-\sqrt{\f{y-x}{4X(x)}}dx \nn
\eea
where
\be
A=-\frac12 \cos\theta d\phi.
\ee
Note that we have judiciously rotated the vielbeine $d\theta$ and $\sin\theta d\phi$. The main reason for the rotation by an angle $2(\tau+\psi)$ is that it eliminates an otherwise cumbersome phase in the associated holomorphic three-form. As a warm up we verify that the above space has $SU(3)$ structure. It, of course, has $SU(3)$ holonomy but here we introduce some notation as well to make contact with the established literature.

Let us define the following 3- and 2-forms $\Omega$ and $J$
\bea
\label{eq:OmegaJ-forms}
\Omega &=&(e^1 + i  e^2 )\wedge(e^4 +i  e^5 )\wedge (e^6 + i e^3), \nonumber \\
J &=& e^1\wedge e^2  + e^4\wedge  e^5  + e^6\wedge e^3
\eea
The main comments is that the above forms satisfy the following $SU(3)$ algebraic constraints

\be
\Omega\wedge J=0, \qquad   \Omega\wedge\bar{\Omega}=- \frac{4}{3}i J\wedge J \wedge J.
\ee
As well as the following differential constraints:

\be
d\Omega=0, \quad  dJ=0, \qquad  d\left(J\wedge J\right)=0.
\ee
Although the last differential constraint follows from dJ=0, these constraints parallel the most general case which we discuss in forthcoming sections. From the resolved cone over $Y^{p,q}$ one can recover the metric on the cone over $Y^{p,q}$  by taking the $x\to -\infty$ limit  as explained in \cite{Martelli:2007pv,Martelli:2007mk}. Introducing
\be
x=-\frac{2}{3}r^2,
\ee
and expanding the metric in the large $r$ limit one finds that the  leading terms in the metric become
\bea
\label{uv}
ds^2&=&dr^2 +\frac{2}{3}r^2\bigg[\frac{1}{4Y(y)}dy^2 +Y(y)(d\psi-\cos\theta d\phi)^2\\\nonumber
& +&\frac14(1-y) (d\theta^2 +\sin^2\theta d\phi^2) +\frac23 (d\tau +(1-y)(d\psi -\cos\theta d\phi))^2 \bigg],
\eea
which is precisely the cone over $Y^{p,q}$. The difference between the above metric and the one presented in equation (\ref{cone-original})  has been explained in various papers \cite{Gauntlett:2004yd,Gauntlett:2004hh} and more generally section 3 of \cite{Martelli:2007pv}. The presentation of equation (\ref{uv}) makes clear the local structure of $Y^{p,q}$ as a $U(1)$ bundle over a K\"ahler-Einstein base. More precisely, the function $Y(y)$ here is proportional to the product $w(y)q(y)$ of the functions defined in (\ref{cone-original}).

\section{Probe analysis}\label{sec:probe}

The question we pose in this section is the following: Is there a probe solution corresponding to a supersymmetric D5 on the resolved
cone over $Y^{p,q}$ such that the backreacted solution corresponds to stacking a large number of such supersymmetric solutions and taking its backreaction into account?

As far as we are aware, this question has not been answered explicitly even in the simpler case of the the conifold, in which
case it is purportedly related to the MN \cite{Maldacena:2000yy} solution. The obvious reason being the existence of the full backreacted solution. We will revisit this question and try to elucidate the situation starting
from the simplest cases which we present explicitly in appendix \ref{app:D5Conifold}.

Probe branes on spaces of the form $AdS_5\times X^5$ where $X^5$ is a Sasaki-Einstein manifold have been systematically studied, for example, the case $T^{1,1}$ was addressed in \cite{Arean:2004mm}, $Y^{p,q}$ in \cite{Canoura2006} and $L^{p,q,r}$ in \cite{Canoura:2006es}. These studies have clarified many aspects, including the possibility of generalizations of these geometries of the form $AdS_5\times X^5$ to cascading regimes and beyond. We will, naturally, build on those works. However, those spaces can be thought as the spaces resulting by taking into consideration the backreaction of D3 branes with the subsequent Maldacena limit. The task at hand for us is simpler as we are concerned with non-backreacted geometries of the form $\mathbb{R}^{1,3}\times CY$ where we consider just $D5$ branes embeddings.

\subsection{Kappa symmetry and supersymmetric branes}

Let us briefly review the formalism of $\kappa$-symmetry used to determine the supersymmetry of a given Dp brane.
We will consider embeddings of D5 branes on $\mathbb{R}^{1,3}\times \check{C}(Y^{p,q})$ which is a
super-symmetric solution to the string equations of motion by virtue of $\check{C}(Y^{p,q})$ being Calabi-Yau. We consider  $\xi^{\mu}$ ($\mu=0,\cdots, 5$)
as a set of worldvolume coordinates and $X^M$ denote ten-dimensional coordinates, the embedding of the brane probe in the background geometry
will be characterized by the set of functions $X^M(\xi^{\mu})$, from which
the induced metric on the world volume is determined as:
\be
g_{\mu\nu}\,=\,\partial_{\mu} X^{M}\,\partial_{\nu} X^{N}\,G_{MN}\,\,,
\label{inducedmetric}
\ee
where $G_{MN}$ is the ten-dimensional metric. Let $e^{\underline{M}}$ be
the frame one-forms of the ten-dimensional metric. These one-forms can be
written in terms of the differentials of the coordinates by means of the
coefficients $E_{N}^{\underline{M}}$:
\beq
e^{\underline{M}}\,=\,E_{N}^{\underline{M}}\,dX^N\,\,.
\eeq
From the $E_{N}^{\underline{M}}$'s and the embedding functions
$X^M(\xi^{\mu})$ we define the induced Dirac matrices on the worldvolume as:
\beq
\gamma_{\mu}\,=\,\partial_{\mu}\,X^{M}\,E_{M}^{\underline{N}}\,\,
\Gamma_{\underline{N}}\,\,,
\label{wvgamma}
\eeq
where $\Gamma_{\underline{N}}$ are constant ten-dimensional Dirac matrices.

The supersymmetric embeddings of the brane probes are obtained by imposing
the kappa-symmetry condition:
\beq
\Gamma_{\kappa}\,\epsilon\,=\,\epsilon\,\,,
\label{kappacondition}
\eeq
where $\epsilon$ is a Killing spinor of the background and $\Gamma_{\kappa}$
is a matrix that depends on the embedding. In order to write the expression
of $\Gamma_{\kappa}$ for the type IIB theory it is convenient to decompose
the complex spinor $\epsilon$ in its real and imaginary parts, $\epsilon_1$
and $\epsilon_2$. These are Majorana--Weyl spinors. They can be subsequently arranged as a two-dimensional vector
\beq
\epsilon=\epsilon_1+i\epsilon_2 ~\longleftrightarrow~ \epsilon =\left(
\begin{array}{c}
\epsilon_1\\
\epsilon_2
\end{array}\right)
\,\,.
\eeq
The dictionary to go from complex to real spinors is:
\beq
\epsilon^* ~\longleftrightarrow~ \tau_3\,\epsilon\,\,,
\,\,\,\,\,\,\,\,\,\,\,\,\,\,\,\,\,\,\,
i\epsilon^* ~\longleftrightarrow~ \tau_1\,\epsilon\,\,,
\,\,\,\,\,\,\,\,\,\,\,\,\,\,\,\,\,\,\,
i\epsilon ~\longleftrightarrow~ -i\tau_2\,\epsilon\,\,,
\label{rule}
\eeq
where the $\tau_i$ ($i=1,2,3$) are the Pauli matrices. If there
are no worldvolume gauge fields on the D5-brane, the kappa symmetry matrix
 is given by \cite{Bergshoeff:1996tu,Bergshoeff:1997kr}:
\beq
\Gamma_{\kappa}\,\epsilon\,=\,{i\over 6!\, \sqrt{-g}}\,\,
\epsilon^{\mu_1\cdots\mu_6}\,\gamma_{\mu_1\cdots\mu_6}\,\epsilon^*
\label{Gammakappad5}\,\,,
\eeq
where $g$ is the determinant of the induced metric $g_{\mu\nu}$ and
$\gamma_{\mu_1\cdots\mu_{6}}$ denotes the antisymmetrized product of
the induced Dirac matrices (\ref{wvgamma}). A more general account of kappa symmetry and calibrations can be found in \cite{Martucci:2005ht,Martucci:2006ij}

The kappa symmetry condition imposes
a new projection on the Killing spinor $\epsilon$ which, in general, will
not be compatible with those already satisfied by $\epsilon$. This is so because the new projections
involve matrices which do not commute with other projections imposed on the spinor.  The only way of making these two conditions
consistent with each other is by requiring the vanishing of the
coefficients of those non-commuting matrices, which will give rise to a
set of first-order BPS differential equations.

The appearance of complex conjugation on the kappa symmetry equation is crucial in what follows as complex conjugation
does not commute with the typical projections imposed on the spinor.

\subsection{Killing spinor for resolved cone $\check{C}(Y^{p,q})$}\label{Sec:KillingSpinor}

In this subsection we first compute the Killing spinor $\epsilon$ in the resolved cone over $Y^{p,q}$. The  metric of the resolved cone over $Y^{p,q}$ was written in equation (\ref{CYM}). Here, for convenience,  we will introduce a slightly different notation
\bea
\eta &=&d\psi-\f{1}{2}\cos \t d\phi.
\eea
More importantly, in this section we consider a simpler sechsbein that is not rotated, namely\footnote{We hope that the use of a different Sechsbein does not confuse the reader as it is used only in this section, next section uses the Sechbein introduced in equation (\ref{CYM}).}
\bea
e^1&=&\f{\sqrt{(1-x)(1-y)}}{2}d\t, \quad e^2=\f{\sqrt{(1-x)(1-y)}}{2}\sin\t d\p\nn\\
e^3&=&\sqrt{\f{X(x)}{(y-x)}}\{d\tau +(1-y)\eta\}, \quad e^4=-\sqrt{\f{y-x}{4Y(y)}}dy\nn\\
e^5&=&\sqrt{\f{Y(y)}{(y-x)}}\{d\tau+(1-x)\eta\}, \quad e^6=-\sqrt{\f{y-x}{4X(x)}}dx,
\eea
To write the spin connection, we use the notation $\hX=\sqrt{\f{X(x)}{y-x}}$, $\hY=\sqrt{\f{Y(y)}{y-x}}$ and $S=\sqrt{(1-x)(1-y)}$.
The Killing Spinor equation is
\be
D_M\ep=\pa _{M}\ep+\f{1}{4}\o _{ab\;M}\G ^{ab}\ep=0.
\ee
We will use the following relations
\be
X'=2x+\f{X}{1-x},\qquad  Y'=-2y+\f{Y}{1-y}.
\ee
It is also convenient to introduce the following projections
\be
P^{12}=\f{1}{2}(1-\G ^{3456}),{\hs{1 cm}}P^{36}=\f{1}{2}(1+\G ^{1245},){\hs{1 cm}}
P^{45}=\f{1}{2}(1-\G^{1236}).
\ee
The Killing spinor equation can be written (see appendix \ref{app:spinconnection} for the explicit expression for the spin connection)
\bea
\pa _{\t}\ep+\f{S}{2}\lf[\f{\hY}{(1-y)}\G ^{14}P^{36}+\f{\hX}{(1-x)}\G ^{16}P^{45}\ri ]\ep&=&0\nn\\
\pa _{\p}\ep-\f{S\sin\t}{2}\lf [\f{\hX}{(1-x)}\G^{13}P^{45}+\f{\hY}{(1-y)}\G^{15}P^{36}\ri ]\ep-\f{\sqrt{XY}\cos\t}{2}\G^{34}P^{12}\ep&&\nn\\
+\f{\cos\t}{2(x-y)^2}\{X(1-y)+Y(1-x)\}\G^{36}P^{12}\ep+\cos\t\G ^{12}P^{45}&&\nn\\
+\f{\cos\t}{2(y-x)}\lf [\lf (\f{1-y}{1-x}\ri )X\G^{12}P^{45}+\lf (\f{1-x}{1-y}\ri )Y\G ^{12}P^{36}+2y(1-x)\G^{36}P^{12} \ri ]\ep&=&0\nn\\
\pa _{\psi}\ep+\G^{36}\ep+\f{\sqrt{XY}}{(y-x)}\G^{34}P^{12}\ep-\f{1}{(x-y)^2}\{X(1-y)+Y(1-x)\}\G^{36}P^{12}\ep&&\nn\\
\f{1}{(x-y)}\lf [\lf (\f{1-y}{1-x}\ri )X\G^{12}P^{45}+\lf (\f{1-x}{1-y}\ri )Y\G ^{12}P^{36}-2y(1-x)\G^{36}P^{12} \ri ]\ep&=&0\nn\\
\pa _{\tau}\ep+\G ^{36}\ep +\f{2y}{x-y}\G^{36}P^{12}\ep+\f{\G^{12}}{(x-y)}\lf [\f{XP^{45}}{(1-x)}+\f{YP^{36}}{(1-y)}\ri ]\ep&&\nn\\
-\f{(X+Y)}{(x-y)^2}\G^{36}P^{12}\ep&=&0\nn\\
\pa _x\ep+\f{1}{2(y-x)}\sqrt{\f{Y}{X}}\G^{35}P^{12}\ep&=&0\nn\\
\pa _y\ep +\f{1}{2(y-x)}\sqrt{\f{X}{Y}}\G^{35}P^{12}\ep&=&0\nn\\
\eea
The three projections $P^{12}, P^{36}$ and $P^{45}$ are not independent. Indeed, they are related as
\be
P^{12}-P^{36}=\G^{1245}P^{45}.
\ee
The equations simplifies considerably if we impose condition
\be
\label{eq:Killcond-resolved}
P^{36}\ep=P^{45}\ep=0.
\ee
The solution for the Killing spinor will be
\be
\label{Killsol}
\ep =e^{-\G ^{36}(\tau+\psi)}P_-^{36}P_+^{45}\ep _0,
\ee
where $\ep _0$ is an arbitrary constant spinor, and
\be
P_-^{36}=\f{1}{2}(1-\G ^{1245}),{\hs{1 cm}}
P_+^{45}=\f{1}{2}(1+\G^{1236}).
\ee
Note that $\G^{36}$ commutes with $P^{36}$ and $P^{45}$ and, moreover, we one can verify that $P_-^{36}P^{36}=P_+^{45}P^{45}=0$. As explained before, the
phase in the spinor is correlated with the fact that the vielbein used here are not rotated by an angle in $2(\tau+\psi)$ as done in
section (\ref{Sec:Review}). We have thus constructed the covariantly constant spinor which determines which embeddings can be supersymmetric.

\subsection{D5 probe in resolved cone  $\check{C}(Y^{p,q})$ geometry}
The ten-dimensional background has the following metric
\bea
ds^2=dx_{1,3}^2+ds_6^2,
\eea
where $ds_6^2$ is the the metric of resolved cone  $\check{C}(Y^{p,q})$ (\ref{CYM}). We consider a D5 probe on this background with embedding coordinates
\be
\xi^\mu=\{x_0,\; x_1,\;x_2,\;x_3,\;\t,\;\p\}
\ee
we take $\tau$ and $\psi$ to be constants and $x$ and $y$ be both functions of $\t$ and $\phi$. The induced gamma matrices are
\bea
\label{eq:inducedgamma}
\g_{x_i}&=&\G_{x_i},\nn\\
\g_{\t}&=&\f{S}{2}\G_1-\f{1}{2}\lf(\f{y_{\t}}{\hY}\G_4+\f{x_{\t}}{\hX}\G_6\ri ), \\
\g_{\p}&=&\f{S}{2}\sin\t \G_2-\f{\cos\t}{2}\{\hX(1-y)\G_3+\hY(1-x)\G_5\}
-\f{1}{2}\lf(\f{y_{\t}}{\hY}\G_4+\f{x_{\t}}{\hX}\G_6\ri ), \nn
\eea
where for example $x_\theta= \frac{\partial x}{\partial \theta}$, and  $\hX=\sqrt{\f{X(x)}{y-x}}$, $\hY=\sqrt{\f{Y(y)}{y-x}}$, $S=\sqrt{(1-x)(1-y)}$.
For the embedding to be supersymmetric, we need to satisfy the kappa symmetry equation
\be
\label{kapypq}
\f{i}{\sqrt{-g}}\g_{x_0x_1x_2x_3\t\p}\ep ^*=\ep .
\ee
From the above expressions in equation (\ref{eq:inducedgamma}) we obtain

\bea
\g _{\t\p}&=&\f{S^2}{4}\sin\t\G_{12}-\f{S\cos\t}{4}[\hat{X}(1-y)\G_{13}+\hat{Y}(1-x)\G_{15}]\nn\\
&-&\f{S}{4}\lf (\f{x_{\p}}{\hat{X}}\G_{16}+\f{y_{\p}}{\hat{Y}}\G_{14}\ri )
+\f{S}{4}\sin\t\lf (\f{x_{\t}}{\hat{X}}\G_{26}+\f{y_{\t}}{\hat{Y}}\G_{24}\ri )\nn\\
&-&\f{(1-y)\cos\t}{4}\lf [y_{\t}\f{\hat{X}}{\hat{Y}}\G_{34}+x_{\t}\G_{36}\ri ]+\f{(1-x)\cos \t}{4}\lf [y_{\t}\G_{45}-x_{\t}\f{\hat{Y}}{\hat{X}}\G_{56}\ri]\nn\\
&+&\f{1}{4\hat{X}\hat{Y}}(y_{\t}x_{\p}-x_{\t}y_{\p})\G_{46}.
\eea
Recall that the spinor satisfies the following projections
\be
\label{eq:projectionypq}
\G_{12}\ep=\G_{45}\ep
\ee
for simplification.
We next check compatibility of above projection conditions with kappa symmetry equation (\ref{kapypq}). We find that only the $\G_{12}$ term of $\g_{\t\p}$ is compatible with both projection conditions; we obtain the following equations
\bea
\label{eq:d5rypqconditions}
x_{\p}&=&0, \qquad y_{\p}=0, \qquad x_{\t}\tan\t =\f{X(1-y)}{(y-x)}, \nn \\
y_{\t}\tan\t &=&\f{Y(1-x)}{(y-x)}, \qquad \f{\hY}{\hX}(1-x)x_{\t}=\f{\hX}{\hY}(1-y)y_{\t}.
\eea
We check that the last equation is not an independent equation and it is consistent with the two equations above it. Removing the explicit parameter $\t$, we reduce the system of equations to the following implicit equation
\be
y_{x}=\f{(1-x)Y(y)}{(1-y)X(x)}.
\ee
The kappa symmetry equation (\ref{kapypq}) then reduces to
\be
i\G_{x}\G_{12}\ep^*=\s\ep
\ee
where $\G_{x}=\G_{x_0x_1x_2x_3}$ and $\s={\tr{sgn}}(\sin\t)$.
The general spinor (\ref{Killsol}) is constrained by Killing spinor equations to be
\be
\ep=e^{-\G_{36}(\tau+\psi)}\eta.
\ee
where $\eta$ is a constant spinor satisfying projection conditions (\ref{eq:projectionypq}). The chirality condition in 10 dimensions reduces to
\be
\G_{x_0x_1x_2x_3123456}\ep=\ep\;\ra\;\G_{12}\ep=-\G_x\ep .
\ee
It simplifies the kappa condition to be
\be
i\eta ^*=\s\eta.
\ee
If we take $\eta=\eta_R+i\eta _I$, then
\bea
\s=1&\ra &\eta _R=\eta _I\nn\\
\s=-1&\ra &\eta _R=-\eta _I.
\eea
So, kappa symmetry equation can be satisfied.


\subsection{Comments on calibrated 2-cycles on $\check{C}(Y^{p,q})$ }
We are interested in verifying the existence of calibrated cycles for the resolved cone over $Y^{p,q}$. Namely, we look for cycles $\Sigma$  verifying the relation that the induced K\"ahler form is the same as the induced volume form on the two cycle, up to a constant phase
\be\label{calibrationC}
J|_{\S}=e^{i\lambda}{\tr{Vol}}|_{\S}.
\ee
We use the K\"ahler form presented in (\ref{eq:OmegaJ-forms}). Let us first consider the solution obtained using kappa symmetry in the previous section, that is, an embedding given by
\bea\label{kapEmb}
y_{x}&=&\f{(1-x)Y(y)}{(1-y)X(x)}\nn\\
x_{\t}\tan\t &=&\f{X(1-y)}{(y-x)}.
\eea
The K\"ahler form reduces to
\bea
J|_{\S}&=&\f{S^2}{4}\sin\t d\t\w d\p +\f{(1-y)\cos\t}{4}x_{\t}d\t \w d\p+\f{(1-x)\cos\t}{4}y_{\t}d\t\w d\p \nn\\
&=&\f{1}{4}\lf [(1-x)(1-y)+\frac{X(1-y)^2+Y(1-x)^2}{(y-x)\tan^2\theta}\ri ]\sin\t d\t\w d\p .
\eea
The induced metric can be simplified to give
\be
ds^2_{\Sigma}=\frac{1}{4}\left[(1-x)(1-y)+\frac{X(1-y)^2+Y(1-x)^2}{(y-x)\tan^2\theta}\right](d\theta ^2+\sin ^2\theta d \phi ^2),
\ee
which results in
\be
{\tr{vol}}|_{\S}=\f{1}{4}\lf [(1-x)(1-y)+\frac{X(1-y)^2+Y(1-x)^2}{(y-x)\tan^2\theta}\ri ]\sin\t d\t\w d\p .
\ee
Hence, the condition (\ref{calibrationC}) is satisfied for our embedding and the two cycle is calibrated in our case.

Given the coordinate parametrization of $\mathbb{CP}^1$ , one might naively consider a 2-cycle  $\Sigma$ defined by the coordinates $(\theta, \phi)$ and all other coordinates constant. Then
\be
J_{|\Sigma}=\frac{(1-x)(1-y)}{4} \sin\theta d\theta \wedge d\phi.
\ee
The induced metric is
\be
ds^2_{\Sigma}=\frac{(1-x)(1-y)}{4}\left(d\theta^2 +\sin^2\theta d\phi^2 \right)+\frac{X(x)}{y-x}(1-y)^2\cos^2 \theta d\phi^2
+ \frac{Y(y)}{y-x}(1-x)^2\cos^2 \theta d\phi^2,
\ee
which results in
\be
vol_{\Sigma}=  \frac{(1-x)(1-y)}{4} \sin\theta d\theta \wedge d\phi \sqrt{1+\cot^2 \theta \left(\frac{X(x)(1-y)}{(y-x)(1-x)} +
\frac{Y(y)(1-x)}{(y-x)(1-y)}\right)}.
\ee
The condition
\be
J_{|\Sigma}=vol_{\Sigma},
\ee
then would require that $X(x)(1-y)^2+Y(y)(1-x)^2=0$.  However, as one can see from (\ref{eq2.7}), this condition can never be obtained in the requisite range of coordinates but approaches calibration as $x\rightarrow x_{-}=y_1$ and as $y\rightarrow y_2$.

\section{Toward NS5-branes in the Resolved Cone over $Y^{p,q}$}\label{sec:NS5sol}

\subsection{Approach through calibration}

As we explained in the introduction, there have been some attempts at the construction of cascading theories using D3 and D5 branes on the cone over $Y^{p,q}$  (\cite{Herzog:2004tr,Burrington:2005zd}). In this manuscript we consider  NS5 branes wrapping a two-cycle in a resolution of the cone over $Y^{p,q}$ .  The geometry of the solution we
seek is non-K\"ahler and  can  be characterized in terms of a real  two-form $J$ and a complex three-form $\Omega$ defining
the $SU(3)$ structure. Demanding supersymmetry imposes certain requirements on these forms. These constraints were derived in \cite{Strominger1986}, and can be written as calibrating conditions \cite{Gauntlett2004},

\bea\label{eq:calibrating}
d\left(e^{-2\phi}\Omega\right)&=&0, \quad e^{2\phi}d\left(e^{-2\phi}J\right)=-\star_6 H_3, \quad d\left(e^{-2\phi}J\wedge J\right)=0.
\eea

In order to guarantee SU(3) structure,  $\Omega$ and $J$ have to satisfy two algebraic constraints,
\bea \label{OO}
\O \w\bar{\O}&=&-\frac{4i}{3}J^3,\qquad  J\w \O=0.
 \eea

One substantially difficult technical problem is the fact that supergravity solutions built on
the cone over $Y^{p,q}$ naturally lead to partial differential equations (PDE). The simplest such example can be seen in the
background with fractional D3 branes of \cite{Herzog:2004tr} where the warp factor is a function of two coordinates $r$ and $y$. A
further attempt to find the chiral symmetry broken phase of the solution runs against similar problems \cite{Burrington:2005zd}. However, in \cite{Herzog:2004tr} and \cite{Burrington:2005zd}
there is  a factorization at play and the solutions admit a relatively simple form.
 One of the most daunting tasks in our case is the fact that for
the resolved cone over $Y^{p,q}$ there is an explicit symmetry between the radial direction $x$ and the angular direction $y$ and no factorization seems possible.


\subsection{NS5 branes wrapping 2 cycle on the resolved cone over $Y^{p,q}$ }\label{sec:resolved}

Consider the following string frame metric:

\be\label{eq:ansatz-resolved}
ds_{str}^2 = dx_{1,3}^2 + e^ {2 g_1} e_1^2 + e^ {2 g_2} e_2^2 + e^ {2 h_1} e_4^2 + e^ {2 h_2} e_5^2 + e^ {2 k_1} e_3^2 + e^ {2 k_2} e_6^2,
\ee
where we have used the sechsbein defined in \eqref{eq:Sechsbein}. The deformation factors
depend on two variables,  $g_1\equiv g_1(x,y),\;g_2\equiv g_2(x,y), \; h_1\equiv h_1(x,y),\;  etc.$  but  we will not write the explicit $(x,y)$ dependence unless needed.

The calibrating conditions only guarantee supersymmetry, we need to supplement them with the Bianchi
identity to ensure that our background is a solution of the IIB equations of motion.
A natural starting point for $H_3$ is,
\be
H_3=\left(F_1(x,y) e_1\wedge e_2   +  F_2(x,y)e_4 \wedge e_5\right)\wedge e_3.
\ee
This ansatz satisfies the asymptotic form of the flux that we expect, that is, it is proportional to the volume form of the topological $S^3$ in the uv. The Bianchi identity
\be
d H_3 =0,
\ee
leads to
\be\label{eq:H3-resolved}
H_3=\frac{1}{(-1 + y)^2 \sqrt{X(x)} }\left(\frac{\sqrt{ y-x}}{(-1 + x) } e_1\wedge e_2   + \frac{1}{\sqrt{y-x}} e_4 \wedge e_5\right)\wedge e_3.
\ee
It can also be verified that this ansatz for $H_3$ is smooth. Imposing the calibrating conditions \eqref{eq:calibrating} on the ansatz given by  \eqref{eq:ansatz-resolved}, \eqref{eq:H3-resolved} and   demanding  integrability   we obtain a system of 11 PDE's  plus two algebraic constraints. The $x$ derivatives equations are,
\be\label{eq:x-deriv-resolved}
\begin{split}
\phi ' &= \Phi[g_1,g_2, k_1, k_2, h_1, h_2, \phi](x,y),  \\
g_i'(x,y)&= G_i[g_1,g_2, k_1, k_2, h_1, h_2, \phi](x,y),\\
h_1'(x,y)&=h_2'(x,y)=H[g_1,g_2, k_1, k_2, h_1, h_2, \phi](x,y),\\
k_i'(x,y)&=K_i[g_1,g_2, k_1, k_2, h_1, h_2, \phi](x,y).\\
\end{split}
\ee
The $y$ derivatives are
\be\label{eq:y-deriv-resolved}
\begin{split}
\dot \phi &= 0,\\
\dot g_i(x,y)&=\widetilde G_i[g_1,g_2, k_1, k_2, h_1, h_2](x,y),\\
\dot h_i(x,y)&=\widetilde H_i [g_1,g_2, k_1, k_2, h_1, h_2](x,y),\\
\dot k_1(x,y)&=\dot k_2(x,y) =\widetilde K[k_1, k_2, h_1, h_2](x,y).
\end{split}
\ee
The algebraic constraints are  given by
\be\label{eq:constraints-resolved}
{\mathcal D}_1[g_1,g_2, k_1, k_2, h_1, h_2](x,y) =0, \qquad \qquad {\mathcal D}_2[g_1,g_2, k_1, k_2, h_1, h_2](x,y) =0.
\ee
In the above expressions $i={1,2}$ and  $K[f_1, f_2...](x,y)$ denotes a functional
of $f_1, f_2.....$ evaluated at the point $(x,y)$ . The explicit form of the equations is given in Appendix \ref{app:eq-resolved}. It is worth emphasizing  that some of the equations in \eqref{eq:x-deriv-resolved},\eqref{eq:y-deriv-resolved},\eqref{eq:constraints-resolved} come from demanding integrability,
$\partial_x \partial_y =\partial_y \partial_x $, and thus ensure that the system is consistent. This system of PDEs together with \eqref{eq:H3-resolved} completely specify the background we are looking for   and constitutes one of our main results.
Let us comment on some features of these equations. The dilaton is always independent of $y$. Thus,  if we consider the exponential of the dilaton to be  related to the strong coupling scale as proposed in
\cite{Maldacena:2000yy} and \cite{Hori2002}
\be
E\sim e^{-\phi}
\ee
then, remarkably, despite the complicated system of PDE's the energy scale is only $r$ dependent. At present, we have not been able to find a closed solution to the system \eqref{eq:x-deriv-resolved},\eqref{eq:y-deriv-resolved},\eqref{eq:constraints-resolved}, we do not see any factorization possible and, most probably, the general solution has to be found numerically.

\subsection{The UV limit:  NS5 wrapping 2 cycle of the cone over $Y^{p,q}$. }
We are interested in  the UV limit $(x\to -\infty)$ of the problem studied in the previous section \ref{sec:resolved}. In this limit,  the leading term of the metric of the resolved cone  is precisely the cone over $Y^{p,q}$, as shown in equation (\ref{uv}).  The $\Omega$ and $J$ of the resolved cone naturally give -in this limit-  the $\Omega$ and $J$ of the cone over $Y^{p,q}$.  Therefore,  the problem  we are after is equivalent to  studying NS5 branes on a 2-cycle of the cone over $Y^{p,q}$.
This limit might be a sort of fixed point of many solutions which differ in the interior (IR); the prototypical examples here would be the KT \cite{Klebanov:2000nc} solution or the singular MN solution
\cite{Maldacena:2000yy}. We start with the following vielbein which is nothing but the $x\to -2r^2/3$ limit of the resolved vielbein (\ref{eq:Sechsbein})

\bea\label{vielbein-uv}
e^1&=&\sqrt{1-y}\,\,(\cos(2(\tau + \psi))d\theta - \sin(2(\tau + \psi)) \sin \theta d\phi), \nonumber \\
e^2&=&\sqrt{1-y}\,\,(\sin(2(\tau + \psi))d\theta +\cos(2(\tau + \psi)) \sin\theta d\phi), \nonumber \\
e^3&=&(d\tau+(1-y)(d\psi + A)), \quad e^4=-\frac{1}{\sqrt{Y(y)}}dy, \nonumber \\
e^5&=&\sqrt{Y(y)}(d\psi + A), \quad e^6=dr,
\eea
such that the $C(Y^{p,q})$ metric \eqref{uv} is written as
\be
ds^2= e_6^2 + r^2 ( \frac{1}{6}(e_1^2 + e_2^2 + e_4^2 ) + \frac{2}{3}e_5^2  + \frac{4}{3} e_3^2)
\ee
One can verify explicitly that the above sechsbein furnishes a pair of $(J, \Omega)$ satisfying all the conditions  for $SU(3)$ structure.

Consider the following ansatz,
\bea\label{eq:ansatz-metric-uv}
ds_{str}^2&=& dx_4^2 +N( e^{g_1}e_1^2 + e^{g_1}e_1^2 +e^{k_1}e_3^2 +e^{h_1}e_4^2 + e^{h_2}e_5^2 + e^{k_2}e_6^2).
\eea
 In the conifold case, one would expect to have $g_1= g_2$. The situation is different for $C(Y^{p,q})$;  it can be shown that  due to the angular dependence $g_1=g_2$ is not a consistent ansatz .

We introduce the following basis,
\begin{align}
\label{Vielbein-uv}
E^1&=e^{g_1}e^1, \quad
E^2=e^{g_2}e^2, \quad
E^3=e^{k1}e^3, \nonumber \\
E^4&=e^{h_1}e^4, \quad E_5=e^{h_2}e^5, \quad E^6=e^{k_2}e^6
\end{align}
In terms of \eqref{Vielbein-uv}, the two-form $J$ and three-form $\Omega$ are given by,
\begin{align}\label{OmegaJcone}
 \Omega &= (E_1 + I E_2)\wedge(E_4 + I E_5)\wedge(E_3 + I E_6),\\
J &= E_1\wedge E_2 + E4\wedge E_5 + E_3\wedge E_6.
\end{align}
By construction these forms satisfy the constraints \eqref{OO}. As explained above, our strategy is to impose the calibrating conditions \eqref{eq:calibrating} on
the ansatz given by \eqref{eq:ansatz-metric-uv} to obtain the BPS equations. We also need
to guarantee that $H_3$ satisfies the Bianchi identity.
Thus, we take
\be\label{eq:H3-cone}
H_3= -\frac{1}{(1 - c y)^2} (e_3 \wedge(e_1 \wedge e_2  +  e_4\wedge e_5))
\ee
which is, by construction, closed: $dH_3=0$.

From the calibrating conditions and differentiability requirement we get  the following   $r$ derivatives equations ,
\be\label{eq:x-deriv-cone}
\begin{split}
\phi ' &= \frac{e^{k_2 -k_1 }}{2 (c y-1)^2}  \left(e^{-{g_1} -{g_2} }-e^{-{h_1} -{h_2} }\right), \\
g_1 '  &= g_2 '  = \dfrac{1}{2 (c y-1)^2}  e^{-g_1 -g_2 -{k1}(r
   ,y)+k_2 },\\
h_1 '  &= h_2'    = \dfrac{1}{2}( e^{2 k_1 }-\dfrac{1}{(c y-1)^2})e^{-h_1 -h_2 -k_1
   +k_2 }, \\
 k_1 '  &= e^{k_2 }\left(\dfrac{1}{2}(e^{-g_1 -g_2 }-e^{-h_1 -h_2 })(\frac{e^{-k_1 }}{ (cy-1)^2} -e^{k_1
   }) \right.\cr & \left.+ e^{-k_1 }\cosh ({\textstyle g_1 -g_2 }) \right).\cr
\end{split}
\ee
The equations for the $y$ derivatives,
\be\label{eq:y-deriv-cone}
\begin{split}
\dot \phi & = \dot k_1  = \dot k_2  = 0,\\
\dot g_1 &= 3 y (c y-1)\dfrac{\sinh ({\textstyle g_2  -g_1  }) }{y^2 (2 c
   y-3)+w}
   e^{h_1 -
   h_2 }
+ c \dfrac{
   e^{-g_1 -g_2 +h_1
   +h_2 }+ 1}{2 c y-2}, \\
\dot g_2 &= -3 y (c y-1)\dfrac{{\textstyle \sinh ( g_2  -g_1  ) }}{y^2 (2 c
   y-3)+w}
   e^{h_1 -
   h_2 } + c \dfrac{
   e^{-g_1 -g_2 +h_1
   +h_2 }+ 1}{2 c y-2}, \\
\dot h_2 &= 3 y (c y-1)\dfrac{{\textstyle \cosh ( g_2  -g_1  ) }}{y^2 (2 c
   y-3)+w}
   e^{h_1 -
   h_2 } + c \dfrac{
   e^{-g_1 -g_2 +h_1
   +h_2 }}{2 c y-2} \cr
& \qquad +\frac{c( w + 9y^2) - 4 c^2 y^3 - 6 y}{2(c y -1)( y^2( 2cy -3) +w)},\\
\dot h_1 &= -3 y (c y-1)\dfrac{{\textstyle \cosh ( g_2  -g_1  ) }}{y^2 (2 c
   y-3)+w}
   e^{h_1 -
   h_2 } + 3c \dfrac{
   e^{-g_1 -g_2 +h_1
   +h_2 }}{2 c y-2} \cr
&\qquad + \dfrac{c}{c y -1)} e^{-2(g_1  + g_2 )+ 2(h_1  + h_2 )} +\frac{-4 c^2 y^3-5 c
   w+3 c y^2+6 y}{2 (c y-1) \left(y^2 (2 c
   y-3)+w\right)}
\end{split}
\ee
and two algebraic constraints
\be\label{eq:constraints}
{\mathcal C}_1(r,y)  =0, \qquad \qquad \text{and} \qquad \qquad {\mathcal C}_2(r,y) =0.
\ee
The explicit expression for ${\mathcal C}_1$ and ${\mathcal C}_2$ is given in Appendix \ref{app:cone}.
The system of equations given by \eqref{eq:x-deriv-cone}, \eqref{eq:y-deriv-cone} and \eqref{eq:constraints} is one of our main results. This system defines the background of NS5 branes wrapping a two cycle in the cone over $Y^{p,q}$ in the simplest case
where the flux is given by \eqref{eq:H3-cone}.

\subsubsection{Asymptotics}
\paragraph{\underline{\boldmath $c\rightarrow 0$}}\hfill \break
Note that for $c=0$, the algebraic constraints \ref{eq:constraints} are identically zero and   equations  \ref{eq:x-deriv-cone} and \ref{eq:y-deriv-cone} admit a simple solution given by,
\begin{align}\label{eq:a0primec0}
 k_1\:&= \: h_2 =\:0, \\
 g_2\:&=\:g_1\:=\: \log r/2,\\
k_2\:&=\:\log 1/2 \cr
\phi\:&= \:\dfrac{1}{4}(-r + \log r + C).
\end{align}
which together with the expression for the flux \eqref{eq:H3-cone}
is, as expected,  the singular Maldacena-N\'u\~nez background. We take this consistency check as evidence that our system correctly describes the analog of the singular MN background for $Y^{p,q}$ spaces.

\paragraph{\underline{Far UV, {\boldmath $r \rightarrow \infty$}}}\hfill\break
To understand the asymptotic properties of our solutions it is worth reviewing five branes solutions.
Let us follow the construction of NS5 brane in \cite{Horowitz:1991cd} and it application to the wrapped NS5 of \cite{Maldacena:2000yy}. In the notation of \cite{Horowitz:1991cd} we work in the isotropic coordinates of equation (21) there and take the decoupling limit where we basically drop the $1$ in the warp functions and in the dilaton. For more about the supersymmetric 5-brane see also \cite{Duff:1990wv,Callan:1991dj}.
The NS5 brane in IIB has the following solution
\bea
ds_{str}^2&=& dx_6^2 + N\left(dr^2 + d\Omega_3^2\right), \nonumber \\
e^\phi&=& e^{\phi_0 -r}, \nonumber \\
H_3&=& Nd\Omega_3.
\eea

What we want as in \cite{Maldacena:2000yy},  is a NS5 wrapping an $S^2$ and thus we are really looking for
\be
ds_6^2=dx_4^2 + N e^{2g}d\Omega_2^2.
\ee
Where our $\Omega_2$ is defined by $e_1$ and $e_2$ above. Thus, in the far UV, where the NS5 we are constructing should look like the NS5 above we expect:

\bea
f(r,y)&\to& a_1 r +F_1(r,y), \nonumber \\
H_3&\to & e_3\wedge e_4\wedge e_5, \qquad {\rm with}\qquad dH_3=0, \nonumber \\
g_1&=&g_2 \to \ln r
\eea

\subsection{Comments on more general ans\"atze}

Let us briefly review the structure of solutions in the case of NS5 branes on conifolds. Our aim is to draw some conclusions which might apply to more general Ansatz\"e for NS5 on $Y^{p,q}$ spaces.
 In the case of $NS5$ branes on conifold-like spaces, a general Ansatz for many Maldacena-N\'u\~nez type of solutions is:
\be
ds_{str}^2= dx_{1,3}^2 +e^{2g}((e_1-a_1(r) e_4)^2+ (e_2-a_2(r) e_5)^2) + e^{2h}(e_4^2 + e_5^2) + e^{2k_1}e_3^2 + e^{2k_2} e_6^2)  \nonumber
\ee
and the flux, $H_3$ also involves a rotation of the basis but with a different function:
\be
H_3 = (e_1 -b_1(r) e_4)\wedge (e_2 -b_2 (r) e_5 ) \wedge e_3 + \tilde{H}_3
\ee
where $\tilde{H}_3$ is a piece necessary to satisfy the Bianchi identity, {\it i.e.}  it is computed
using $dH_3=0$.
The solutions can be classified as belonging to one of the following cases,
\be\label{eq:clasification}
\begin{array}{ccc}
 a_1=a_2=0 & b_1=b_2=0 & {\rm Singular \,\,\,\, MN}  \\
 a_1=a_2=a & b_1=b_2=a& {\rm Regular   \,\,\,\, MN}  \\
 a_1=a_2=a & b_1=b_2=b& {\rm Regular   \,\,\,\,MM \, seed}  \\
 a_1, a_2 &  b_1, b_2& {\rm Reduces\, to\, previous, BPS}
\end{array}
\ee
Even for solutions as general as those discussed in  \cite{Casero:2006pt},
the BPS equations force\footnote{We thank Carlos N\'u\~nez for various comments and clarifications on this point.} $a_1=a_2$ and $b_1=b_2$.

For NS5 on the resolved cone over $Y^{p,q}$ more   general Ans\"atze than the one presented here should exist. We believe they will follow a similar classification as the ones on the conifold, that is, they will involve two deformation functions in the metric and two different
functions in the $H_3$.
 However, in our case it is not quite clear whether the BPS equations force a similar relationship among $a_1$ and $a_2$ and between $b_1$ and $b_2$. It is quite possible that the dependence in two coordinates implies different relationships that become those only in the large radius or  conifold limit which should involve large radius asymptotics or $c\to 0$ in the language of the $Y^{p,q}$ metric.

If a classification similar  to \eqref{eq:clasification} holds for NS5 branes on $\check{C}(Y^{pq})$ the solution presented in the present work  corresponds to $a_1=a_2=b_1=b_2=0$.
More  general ans\"atze should exist and are currently under investigation.

Finding more general ansatz\"e naturally leads to a search
for an interpolating solution.
Recall that in \cite{Papadopoulos:2000gj}, Papadopoulos and
Tseytlin proposed a general Ansatz for backgrounds with
$SU(3)$ structure arising from  five branes wrapped on 2-spheres
on the conifold and its resolutions. Using the PT anstaz an
interpolating solution was later built in \cite{Butti:2004pk}.
We can foresee that a similar program can be carried out
for the cone over $Y^{p, q}$. However, the general form for the
complex structure and K\"alher form
presented in \cite{Papadopoulos:2000gj} was obtained
assuming that they depend only on the radial coordinate.
Thus, we first have to revisit the issue of what is the
general Ansatz for $\Omega$ and $J$ for a manifold with $SU(3)$
structure when the complex structure and K\"ahler form
depend not only on $r$ but also on an angular variable, $y$.
It is not {\it a priori} clear to us if the $\Omega$ and $J$ of
\cite{Papadopoulos:2000gj},\cite{Butti:2004pk} are general
enough for this case.

\section{Conclusions and future directions}
In this paper we have discussed the construction of supersymmetric five branes wrapping a 2-cycle in the resolved cone over $Y^{p,q}$. We have studied the problem at probe level and after finding encouraging evidence move on to the full problem. Our main result was presented in section \ref{sec:resolved} where we presented and ansatz and demonstrated its consistency and the fact that some limits are correctly reproduced. This is a first step in what should be a long program toward the full construction and understanding of five branes on the resolved cone $\check{C}(Y^{p,q})$.
In what follows we outline a few interesting problems some of which we would like to tackle in the future.

\noindent
{\it Numerical study of the system:} Given that we understand the uv asymptotic of the system rather well it would be nice to try to use the asymptotics as boundary conditions in the construction of numerical solutions. We were able to successfully generate some of the series analysis that usually precedes such numerical efforts. It is worth noticing that in some limits certain separation of variables seems possible.

\noindent
{\it Generalizing the Ansatz:} The Ansatz that we considered was limited, in the language of table (\ref{eq:clasification}) to the $a=b=0$. It would be useful to consider the more general cases. Along the same lines, and as stated at the end of section \ref{sec:NS5sol}, it is plausible that this generalization of the Ansatz goes hand in hand with a generalization of the $SU(3)$ structure forms.

\noindent
{\it Chain of dualities and generating solution techniques:} The main motivation for our work is the possibility of performing a chain of duality along the lines of
\cite{Maldacena:2009mw} to obtain a background describing D3 and D5 branes. More generally, we expect the cone over $Y^{p,q}$ to provide a version of the brane/flux transition anticipated by Vafa in the context of Calabi-Yau manifolds \cite{Vafa:2000wi}. We established a framework to construct the gravity solution corresponding to fivebranes wrapping the $S^2$ in the resolved cone over $Y^{p,q}$; there is a potential running of the resolution parameter as in the case discussed in \cite{Maldacena:2009mw}. We expect the final solution to have the topology of the ``deformed'' $C(Y^{p,q})$, that is, a solution with an $S^3$ which has finite size at the tip. It is also worth noting that the chain of dualities has recently been reinterpreted and generalized in \cite{Gaillard:2010qg,Minasian:2009rn,Chen:2010bn} and the implications to five branes on $\check{C}(Y^{p,q})$ could be far reaching.

\noindent
{\it The field theory:} We have not discussed the field theory side. Although the baryonic branch seems to be the natural venue, it is worth mentioning  that there is certain universality in the sense discussed in \cite{Minasian:2009rn} where a deformation along the baryonic branch looks more like a symmetry of the supergravity equations. It would be interesting to understand precisely that relationship in this context. Of course, the whole idea of a ``baryonic'' branch is suspect in view of the works  \cite{Bertolini:2005di,Berenstein:2005xa,Franco:2005zu} as we mentioned in the introduction, that is, equivalent to having a supergravity solution build around a conformal Calabi-Yau space.

\noindent
{\it Connection to cascading solutions:}  Another very interesting question is the precise relation of the five brane solution to the cascading backgrounds constructed in \cite{Herzog:2004tr,Burrington:2005zd}.  Simply following the chain of duality presented in
\cite{Maldacena:2009mw} in the opposite direction does not seem to land us in an ansatz similar to our starting point. It could be, as explained nicely in \cite{Butti:2004pk}, that the structure of a conformal Calabi-Yau space exist only perturbatively in the supergravity family of solutions.

\noindent
{\it Flavor:} The addition of backreacted flavors to these solutions is another interesting and active direction. Indeed, recently, supergravity backgrounds dual to flavored field theories have been found in a variety of cases \cite{Casero:2006pt}, \cite{Casero2007}, \cite{Caceres2007},  \cite{HoyosBadajoz:2008fw}, \cite{Bigazzi2008a}, \cite{Bigazzi2008}.

\noindent
{\it Construction of black holes on this background:} More ambitiously, we mention the construction of black hole on this background and on the flavored backgrounds that could be constructed. This is a significantly more difficult endeavor as it forces us to deal directly with the equations of motion since supersymmetry has to be given up. There have been, however, some encouraging results in the context of the conifold \cite{PandoZayas:2006sa,Mahato:2007zm} and of the MN-like  backgrounds with backreacted flavors  \cite{Caceres2009,Bigazzi2009}.

\noindent
{\it Toward NS5 branes on the resolved cone over $L^{p,q,r}$:} Although much about the field theory and the interpretation of probes on $AdS_5\times L^{p,q,r}$ is known, the metric of the resolution of the cone over $L^{p,q,r}$ is not explicitly known. It is possible that the probe approach discussed here could be applied to understand the possibility of constructing a resolution of the cone over  $L^{p,q,r}$, that is, a construction of  $\check{C}(L^{p,q,r})$. Note that in the case of the conifold and of the cone over $Y^{p,q}$, the 2-cycle that gets a finite volume is already present in the unresolved geometry.

\section*{Acknowledgments}
We are grateful to C. N\'u\~nez for various comments and suggestions. We are also thankful to J. Gaillard and A. Ramallo for important clarifications. E.C and L.P-Z. are thankful to the Aspen Center for Physics for hospitality during the initial stages of this project.
E.C. and V.G.J.R. thank the Michigan Center for Theoretical Physics for hospitality at various stages of this project. E.C also thanks the Theory Group at the University of Texas at Austin for hospitality.
This work is  partially supported by Department of Energy under grant DE-FG02-95ER40899 to the University of Michigan, by the National Science Foundation under  Grant No. PHY-0455649, NSF 0652983 to the University of Iowa and by CONACYT's grants  No. 50760 and No.104649.
\appendix

\section{Details of probe calculation}
\subsection{Spin connection for resolved cone over $Y^{p,q}$ }\label{app:spinconnection}

The relevant components of the one form spin connection are
\bea
\o _{12\p}&=&-\cos \t+\f{\cos \t}{2(y-x)}\lf [\f{1-y}{1-x}X+\f{1-x}{1-y}Y\ri ], \quad \o_{12\tau}=\f{1}{(x-y)}\lf [\f{X}{1-x}+\f{Y}{1-y}\ri ]\nn\\
\o _{12\psi}&=&\f{1}{x-y}\lf [\f{1-y}{1-x}X+\f{1-x}{1-y}Y\ri ], \quad \o _{13\p}=-\f{\hX S}{2(1-x)}\sin\t, \quad \o _{14\t}=\f{S\hY}{2(1-y)}\nn\\
\o _{15\phi}&=&-\f{S\hY}{2(1-y)}\sin\t\nn
\eea
\bea
\o_{16\t}&=&
\o _{23\t}=\f{S\hX}{2(1-x)}, \quad \o _{24\p}=\f{S\hY}{2(1-y)}\sin\t\nn\\
\o _{25\t}&=&\f{S\hY}{2(1-y)}, \quad \o _{26\p}=\f{S\hX}{2(1-x)}\sin\t, \quad \o _{34\psi}=\f{\sqrt{XY}}{y-x}
\eea
\bea
\o _{34\p}&=&-\f{\sqrt{XY}}{2(y-x)}\cos\t, \quad \o _{35y}=\f{1}{2(y-x)}\sqrt{\f{X}{Y}}, \quad \o _{35x}=\f{1}{2(y-x)}\sqrt{\f{Y}{X}}\nn\\
\o _{36\tau}&=&\f{1}{(x-y)^2}((x-y)X'-X-Y),\nn\\
 \quad \o _{36\psi}&=&\f{1}{(x-y)^2}[(1-y)\{(x-y)X'-X\}-(1-x)Y)]\nn\\
\o _{36\phi}&=&-\f{\cos\t}{2}\o _{36\psi}, \quad \o _{45\tau}=-\f{1}{(x-y)^2}(X+(x-y)Y'+Y)\nn
\eea
\bea
\o _{45\psi}&=&-\f{1}{(x-y)^2}[-X(1-y)+\{(x-y)Y'+Y\}(1-x)], \quad \o _{45\p}=-\f{\cos\t}{2}\o _{45\psi}\nn\\
\o _{46y}&=&\f{1}{2(x-y)}\sqrt{\f{X}{Y}}, \quad \o _{46x}=\f{1}{2(x-y)}\sqrt{\f{Y}{X}}\nn\\
\o _{56\psi}&=&\hX\hY, \quad \o _{56\p}=-\f{1}{2}\hX\hY\cos\t
\eea

These are the ingredients needed to write the equations for the Killing spinor in section \ref{Sec:KillingSpinor}.

\subsection{D5 probe in conifold geometry}\label{app:D5Conifold}
To build up intuition and for completeness, we also consider this simpler space. Let us consider a D5 probe on $\mathbb{R}^{1,3}\times \rm{Conifold}$. First we determine the covariantly constant spinor using the metric
\bea
ds_{10}^2&=&dx_{3,1}^2+ds_6^2\\
ds_6^2&=&\f{r^2}{6}\lf (d\t _1^2+\sin ^2\t _1 d\p ^2_1+d\t _2^2+\sin ^2\t _2 d\p ^2_2\ri )+\f{r^2}{9}(d\psi +\cos\t_1d\p _1+\cos \t _2d \p _2)^2+dr^2.\nn
\eea
We choose the veilbeins
\bea
e^1&=&\f{r}{\sqrt{6}}d\t _1, \quad e^2=\f{r}{\sqrt{6}}\sin \t _1d\p _1\nn\\
e^3&=&\f{r}{\sqrt{6}}d\t _2, \quad e_4=\f{r}{\sqrt{6}}\sin \t _2d\p _2\nn\\
e^5&=&\f{r}{3}(d\psi +\cos \t_1d\p _1+\cos\t_2d\p _2), \quad e^6=dr
\eea
The spin connections are
\bea
\o _{12}&=&-\f{\sqrt{6}}{r}\cot\t _1e^2+\f{e^5}{r}, \quad \o _{15}=\f{e^2}{r}, \quad \o _{16}= \f{e^1}{r}, \quad  \o _{25}=-\f{e^1}{r}, \quad
\o_{26}=\f{e^2}{r}\nn\\
\o _{34}&=&-\f{\sqrt{6}}{r}\cot\t _1e^4+\f{e^5}{r}, \quad \o _{35}=\f{e^4}{r}, \quad
\o _{36}= \f{e^3}{r}, \quad \o _{45}=-\f{e^3}{r}, \quad \o_{46}=\f{e^4}{r}, \quad \o _{56}=\f{e^5}{r}\nn\\
\eea
The Killing spinor equation
\be
D_{\m}\ep=\pa _{\m}\ep+\f{1}{4}\o _{ab\m}\G ^{ab}\ep=0
\ee
 This equation is simpler than the analogous computations for $AdS_5\times X^5$ presented explicitly in \cite{Arean:2004mm,Canoura2006,Canoura:2006es} since it does not contain the terms coming from the 5-form. However, there are many similarities in the form of the solution. In particular, for the above background the equations lead to only one non-trivial equation, if we consider projections
\be
\label{project}
\G^{12}\ep=\G^{34}\ep.
\ee
The non-trivial equation is
\be
\pa _{\psi}\ep +\f{1}{2}\G^{12}\ep=0.
\ee
So the solution is
\be\label{epsi}
\ep=e^{-\f{1}{2}\G^{12}\psi}\eta
\ee
where $\eta$ is a constant spinor satisfying the projections (\ref{project}).

Next we put a D5 probe in this background and check kappa symmetry. We consider the embedding
\be
\label{embed}
\xi^\mu=\{x_0,\;x_1,\;x_2,\;x_3,\;\t_1=\t,\;\phi_1=\p\}
\ee
with $r,\; \psi$=constant and $\t _2,\;\p _2$ being functions of $\t$ and $\p$.
The kappa symmetry equation is
\be
\label{kapp}
\f{i}{\sqrt{-g}}\g_{x_0x_1x_2x_3\t\p}\ep ^{*}=\ep.
\ee
The induced matrices are
\bea
\g _{x_i}&=&\G _{x_i}\nn\\
\g_{\t}&=&\f{r}{\sqrt{6}}\lf \{\G_1+\pa _{\t}\t _2\G_3+\sin \t_2\pa_{\t}\p _2\G_4\ri \}+\f{r}{3}\cos\t_2\pa _{\t}\p _2\G_5\nn\\
\g_{\p}&=&\f{r}{\sqrt{6}}\lf \{\sin \t_1\G_2+\pa _{\p}\t _2\G_3
+\sin\t _2\pa _{\p}\p _2\G_4\ri \}
+\f{r}{3}\{\cos \t _1+\cos \t_2\pa _{\p}\p _2\}\G_5.
\eea
This leads to
\bea
\g_{\t\p}&=&
\frac{r^2}{6}\sin\theta_1\G_{12}+ \frac{r^2}{6}\partial_\phi\theta_2\G_{13}+\frac{r^2}{6}\sin\theta_2\partial_\phi\phi_2\G_{14}
+ \frac{r^2}{3\sqrt{6}}\left(\cos\theta_1+\cos\theta_2\partial_\phi\phi_2\right)\G_{15}\nonumber \\
&+&\frac{r^2}{6}\sin\theta_1\partial_\theta\theta_2\G_{32}
+\frac{r^2}{6}\sin\theta_2\left(\partial_\theta\theta_2\partial_\phi\phi_2-\partial_\theta\phi_2\partial_\phi\theta_2\right)\G_{34} \nonumber \\
&+&\frac{r^2}{3\sqrt{6}}\left(\cos\theta_1 \partial_\theta\theta_2 +\cos\theta_2(\partial_\phi\phi_2\partial_\theta\theta_2 -\partial_\theta\phi_2\partial_\phi\theta_2)\right)\G_{35} \nonumber \\
&+&\frac{r^2}{6}\sin\theta_1\sin\theta_2\partial_\theta\phi_2\G_{42}+ \frac{r^2}{3\sqrt{6}}\cos\theta_1\sin\theta_2\partial_\theta\phi_2 \G_{45}
+  \frac{r^2}{3\sqrt{6}}\cos\theta_2\sin\theta_1\partial_\theta\phi_2\G_{52}. \nonumber
\eea

We need the kappa symmetry equation to be compatible with the projections equations (\ref{project}). We find that the only surviving terms are proportional to  $\G_{12}$, $\G_{13}$ and $\G_{14}$ in $\g _{\t\p}$ which satisfy this criteria. Eliminating the coefficients of $\G_{13}$ and $\G_{14}$ gives these equations.  Requiring that $\phi_2 = a + b \phi$, with $a$ and $b$ constants, satisfies the expression
\be
\label{eq:cond-conifold-1}
\pa _{\t}\p_2=0,
\ee
as well as guarantees that $\theta_2$ is a function only of $\theta$.  The only equation that needs to be solved is
\bea
\label{eq:cond-conifold-2}
\sin \t_2(\t) b -\sin\t \pa _{\t}\t _2&=&0.\nn
\eea
This leads to the solution
\be \t_2(\t) = 2 \arctan{e^{c} (\cos{\frac{\t}{2}})^{-b} (\sin{\f{\t}{2}})^{b}},
\ee
where $c$ is a constant.  Therefore we write $\gamma_{\t \p}$ as
\be
\g_{\t \p}= \frac{1}{6}\left( \f{ 4 b^2 e^{(2 c)} (\cos{\frac{\t}{2}})^{2b} (\sin{\f{\t}{2}})^{2b}}{(\cos{\frac{\t}{2}})^{2b}  + e^{2 c} (\sin{\f{\t}{2}})^{2b}} + \sin{\t} \right) \G_{12}.
\ee

For $b=-1$ and $c=0$ this gives
\be
\g_{\t \p}= \frac{1}{3}\sin{\t} \,\G_{12},
\ee
with $\t_2 = \pi - \t$, $\phi_2 = - \phi$.

For $b=1$ this gives
\be
\g_{\t \p}= \frac{1}{3}\sin{\t} \,\G_{12},,
\ee
with $\t_2 =  \t$, $\phi_2 =  \phi$.

Note that our analysis shows that the cycle discussed in appendix A of \cite{Herzog:2001xk}: $\theta_2=\theta_1$ and $\phi_2=-\phi_1$ is not supersymmetric.\footnote{We thank A. Ramallo and J. Gaillard for a discussion of this point.}

\subsection{Calibrated 2-cycles on the conifold}
In this section we show the existence of calibrated cycles $\Sigma$ such that
\be
J_{|\Sigma}=vol_{\Sigma}.
\ee
The embeddings we consider are of the form: $r=r(\theta_1, \phi_1), \theta_2=\theta_2(\theta_1, \phi_1), \phi_2=\phi_2(\theta_1, \phi_1), \psi=\psi(\theta_1, \phi_1)$. A particular solution is

\bea
\partial_\theta \theta_2&=&1, \quad \partial_\phi \theta_2 = \frac{5-11\cos\theta_2}{16\sqrt{3}\cos\theta_2}, \nonumber \\
\partial_\theta \psi&=& \sqrt{3}, \quad \partial_\phi \psi=-\frac{2}{\cos\theta_2}, \qquad \partial_\phi r=0, \quad \partial_\theta r=0, \nonumber \\
\partial_\phi \phi_2&=&1, \quad \partial_\theta\phi_2=0.
\eea

Another interesting calibrated cycle is
\bea
\partial_{\theta}\phi&=&0, \qquad \partial_\phi \phi_2=-1, \nonumber \\
\partial_\theta\theta_2&=& 1, \qquad \partial_\phi\theta_2=0, \nonumber \\
\partial_\theta \psi&=& 1, \qquad \partial_\phi \psi=0, \nonumber \\
\partial_\theta r&=& \frac{r\sqrt{1+(1+\sqrt{1+18\csc^2\theta +\csc^4\theta})\sin^2\theta}}{3\sqrt{2}}, \nonumber \\
\partial_\phi r&=&\frac{r\sqrt{1+(1+\sqrt{1+18\csc^2\theta +\csc^4\theta})\sin^2\theta}}{3\sqrt{2}}.
\eea

\section{Equations of motion for NS5 branes wrapping 2-cycle in the resolved cone $\check{C}(Y_{p,q})$}
\label{app:eq-resolved}

In this appendix we present the explicit form of the equations  \eqref{eq:x-deriv-resolved}-\eqref{eq:y-deriv-resolved} and the constraints \eqref{eq:constraints-resolved}.

The first order independent equations obtained from the calibrating conditions are,

\begin{align}
\partial_x k_1(x,y)&= \frac{x
   e^{g_1(x,y)-g_2(x,y)-k_1(x
   ,y)+k_2(x,y)}}{2 X(x)}+\frac{x
   e^{-g_1(x,y)+g_2(x,y)-k_1(
   x,y)+k_2(x,y)}}{2
   X(x)}\cr\cr & +\frac{(x-y)
   e^{-g_1(x,y)-g_2(x,y)-k_1(
   x,y)+k_2(x,y)}}{4 (x-1) (y-1)^2
   X(x)}-\frac{e^{-h_1(x,y)-
   h_2(x,y)-k_1(x,y)+k_2(x,y)}}{4
   (y-1)^2
   X(x)}\cr \cr & +\frac{e^{-g_1(x,y)-
   {g_2}(x,y)+k_1(x,y)+k_2(x,y)}}{2-
   2
   x}-\frac{e^{-h_1(x,y)-h_2(x,y)+ k_1(x,y)+k_2(x,y)}}{2 x-2
   y}\cr \cr & +\frac{(y-x) X'(x)+X(x)}{2
   X(x) (x-y)}
\end{align}

\begin{align}
\partial_x g_1(x,y)&= \frac{(x-y) \phi(x,y)
   e^{-g_1(x,y)-g_2(x,y)-k_1(
   x,y)+k_2(x,y)}}{2 (x-1) (y-1)^2
   X(x) (4 \phi(x,y)+1)}\cr \cr &+\frac{(2
   \phi(x,y)+1)
   e^{-h_1(x,y)-h_2(x,y)-k_1(
   x,y)+k_2(x,y)}}{4 (y-1)^2 X(x)
   (4 \phi(x,y)+1)}-\frac{x
   e^{g_1(x,y)-g_2(x,y)-k_1(x
   ,y)+k_2(x,y)}}{2 X(x)}\cr\cr &+\frac{x
   e^{-g_1(x,y)+g_2(x,y)-k_1(
   x,y)+k_2(x,y)}}{2
   X(x)}+\frac{e^{-g_1(x,y)-
   {g_2}(x,y)+k_1(x,y)+k_2(x,y)}}{2
   (x-1)}+\frac{1}{2-2 x}
\end{align}

\begin{align}
\partial_x g_2(x,y)&= \frac{(x-y) \phi(x,y)
   e^{-g_1(x,y)-g_2(x,y)-k_1(x,y)+k_2(x,y)}}{2 (x-1) (y-1)^2 X(x)(4 \phi(x,y)+1)}\cr\cr &+\frac{(2\phi(x,y)+1)
   e^{-h_1(x,y)-h_2(x,y)-k_1( x,y)+k_2(x,y)}}{4 (y-1)^2 X(x)(4 \phi(x,y)+1)} +
\frac{x e^{g_1(x,y)-g_2(x,y)-k_1(x,y)+k_2(x,y)}}{2 X(x)}\\
&-\frac{x e^{-g_1(x,y)+g_2(x,y)-k_1(x,y)+k_2(x,y)}}{2 X(x)}+\frac{e^{-g_1(x,y)-
   g_2(x,y)+k_1(x,y)+k_2(x,y)}}{2(x-1)}+\frac{1}{2-2 x}
\end{align}

\begin{align}
\partial_x \phi(x,y)= \frac{(x-y) \phi(x,y)
   e^{-g_1(x,y)-g_2(x,y)-k_1(
   x,y)+k_2(x,y)}}{2 (x-1) (y-1)^2
   X(x) (4 \phi(x,y)+1)}-\frac{\phi(x,y)
   e^{-h_1(x,y)-h_2(x,y)-k_1(x,y)+k_2(x,y)}}{2 (y-1)^2 X(x)(4 \phi(x,y)+1)}
\end{align}

\begin{align}
\partial_x h_1(x,y)&= -\frac{(x-y) (2
   \phi(x,y)+1)
   e^{-g_1(x,y)-g_2(x,y)-k_1(
   x,y)+k_2(x,y)}}{4 (x-1) (y-1)^2
   X(x) (4 \phi(x,y)+1)} -\frac{1}{2 x-2 y}\cr\cr &-\frac{\phi(x,y)
   e^{-h_1(x,y)-h_2(x,y)-k_1(
   x,y)+k_2(x,y)}}{2 (y-1)^2 X(x)
   (4
   \phi(x,y)+1)}+\frac{e^{-h_1(x,y)-h_2(x,y)+k_1(x,y)+k_2(x,y)}}{2 x-2
   y}
\end{align}

\begin{align}
\partial_x h_2(x,y)&= -\frac{(x-y) (2
   \phi(x,y)+1)
   e^{-g_1(x,y)-g_2(x,y)-k_1(
   x,y)+k_2(x,y)}}{4 (x-1) (y-1)^2
   X(x) (4 \phi(x,y)+1)}-\frac{1}{2 x-2 y}\cr\cr &-\frac{\phi(x,y)
   e^{-h_1(x,y)-h_2(x,y)-k_1(
   x,y)+k_2(x,y)}}{2 (y-1)^2 X(x)
   (4
   \phi(x,y)+1)}+\frac{e^{-h_1(x,y)-h_2(x,y)+k_1(x,y)+k_2(x,y)}}{2 x-2
   y}
\end{align}

\begin{align}
\partial_y h_2(x,y)&= -\frac{y
   e^{g_1(x,y)-g_2(x,y)+h_1(x
   ,y)-h_2(x,y)}}{2 Y(y)}-\frac{y
   e^{-g_1(x,y)+g_2(x,y)+h_1(
   x,y)-h_2(x,y)}}{2
   Y(y)}\cr\cr &+\frac{e^{-g_1(x,y)-g_2(x,y)+h_1(x,y)+h_2(x,y)}}{2-
   2
   y}+\frac{e^{h_1(x,y)+h_2(x,y)-k_1(x,y)-k_2(x,y)}}{2 x-2
   y}\cr\cr &-\frac{(x-y) Y'(y)+Y(y)}{2
   Y(y) (x-y)}
\end{align}

\begin{align}
\partial_y g_1(x,y)&= \frac{y
   e^{g_1(x,y)-g_2(x,y)+h_1(x
   ,y)-h_2(x,y)}}{2 Y(y)}-\frac{y
   e^{-g_1(x,y)+g_2(x,y)+h_1(
   x,y)-h_2(x,y)}}{2
   Y(y)}\cr\cr &+\frac{e^{-g_1(x,y)-g_2(x,y)+h_1(x,y)+h_2(x,y)}}{2
   (y-1)}+\frac{1}{2-2 y}
\end{align}

\begin{align}
\partial_y g_2(x,y)&= -\frac{y
   e^{g_1(x,y)-g_2(x,y)+h_1(x
   ,y)-h_2(x,y)}}{2 Y(y)}+\frac{y
   e^{-g_1(x,y)+g_2(x,y)+h_1(
   x,y)-h_2(x,y)}}{2
   Y(y)}\cr\cr &+\frac{e^{-g_1(x,y)-g_2(x,y)+h_1(x,y)+h_2(x,y)}}{2
   (y-1)}+\frac{1}{2-2 y}
\end{align}

\begin{align}
\partial_y \phi(x,y)= 0
\end{align}

\begin{align}
\partial_y k_1(x,y)=
   \frac{e^{h_1(x,y)+h_2(x,y)-k_1(x,y)-k_2(x,y)}}{2 y-2
   x}+\frac{1}{2 x-2 y}
\end{align}

\begin{align}
\partial_y k_2(x,y)=
   \frac{e^{h_1(x,y)+h_2(x,y)-k_1(x,y)-k_2(x,y)}}{2 y-2
   x}+\frac{1}{2 x-2 y}.
\end{align}

Furthermore,  demanding $\partial_x\partial_y =\partial_y\partial_x$  gives two more equations,
\begin{align}
\partial_y h_1(x,y)&=
   \frac{(x-y) \exp (-2 g_1(x,y)-2
   g_2(x,y)+2 h_1(x,y)+2
   h_2(x,y))}{(x-1) (y-1)}\cr\cr &+\frac{y
   e^{g_1(x,y)-g_2(x,y)+h_1(x
   ,y)-h_2(x,y)}}{2 Y(y)}+\frac{y
   e^{-g_1(x,y)+g_2(x,y)+h_1(
   x,y)-h_2(x,y)}}{2
   Y(y)}\cr \cr &+\frac{3
   e^{-g_1(x,y)-g_2(x,y)+h_1(
   x,y)+h_2(x,y)}}{2
   (y-1)}-\frac{e^{h_1(x,y)+h_2(x,y
   )-k_1(x,y)-k_2(x,y)}}{2 x-2
   y}\cr\cr & +\frac{(y-1) (x-y)
   Y'(y)+Y(y) (-4 x+5 y-1)}{2
   (y-1) Y(y) (x-y)} \nonumber
\end{align}
\begin{align}
\partial_x k_2(x,y)&= \frac{x y
   (x-y) \left(e^{2 g_1(x,y)}-e^{2
   g_2(x,y)}\right)^2 \exp (-2
   (g_1(x,y)+g_2(x,y))-2
   h_2(x,y)+2
   k_2(x,y))}{X(x)
   Y(y)}\cr\cr &+\frac{2 (x-y) \exp (-2
   g_1(x,y)-2 g_2(x,y)+2
   k_1(x,y)+2 k_2(x,y))}{(x-1)
   (y-1)}\cr\cr& -\frac{2 (x-1) \exp (-2
   h_1(x,y)-2 h_2(x,y)+2
   k_1(x,y)+2 k_2(x,y))}{(y-1)
   (x-y)}\cr\cr &-\frac{(x-y) (8 \phi(x,y)+3)
   e^{-g_1(x,y)-g_2(x,y)-k_1(
   x,y)+k_2(x,y)}}{4 (x-1) (y-1)^2
   X(x) (4
   \phi(x,y)+1)}\cr\cr &+\frac{e^{-h_1(x,y)-h_2(x,y)-k_1(x,y)+k_2(x,y)}}{4
   (y-1)^2 X(x) (4 \phi(x,y)+1)}-\frac{x
   e^{g_1(x,y)-g_2(x,y)-k_1(x
   ,y)+k_2(x,y)}}{2 X(x)}\cr\cr&-\frac{x
   e^{-g_1(x,y)+g_2(x,y)-k_1(
   x,y)+k_2(x,y)}}{2
   X(x)}+\frac{3
   e^{-g_1(x,y)-g_2(x,y)+k_1(
   x,y)+k_2(x,y)}}{2 (x-1)}\cr\cr&+\frac{9
   e^{-h_1(x,y)-h_2(x,y)+k_1(
   x,y)+k_2(x,y)}}{2 (x-y)}+\frac{(x-y)
   X'(x)-5 X(x)}{2 X(x)
   (x-y)}  \nonumber
   \end{align}

and  two constraints given by;
\begin{align}
&\mathcal{D}_1 = -\frac{(x-y)^2 \exp (-3 g_1(x,y)-3
   g_2(x,y)+2 h_1(x,y)+2
   h_2(x,y)-k_1(x,y)+k_2(x,y)
   )}{(x-1)^2 (y-1)^3 X(x)}\cr\cr&+\frac{2
   (y-x) \exp (-2 g_1(x,y)-2
   g_2(x,y)+h_1(x,y)+h_2(x,y)
   -k_1(x,y)+k_2(x,y))}{(x-1)
   (y-1)^3 X(x)}\\&-\frac{2 (x-y) \exp (-3
   g_1(x,y)-3 g_2(x,y)+2
   h_1(x,y)+2
   h_2(x,y)+k_1(x,y)+k_2(x,y)
   )}{(x-1)^2 (y-1)}\cr\cr&+\frac{2 \exp (-2
   g_1(x,y)-2
   g_2(x,y)+h_1(x,y)+h_2(x,y)
   +k_1(x,y)+k_2(x,y))}{(x-1)
   (y-1)}\cr\cr&-\frac{\exp (-2 g_1(x,y)-2
   g_2(x,y)+2 h_1(x,y)+2
   h_2(x,y))}{(x-1)^2}-\frac{e^{-g_1(x,y)-g_2(x,y)-k_1(x,y)+k_2(x,y)}}{(y-1)^3
   X(x)}\cr\cr&+\frac{e^{-g_1(x,y)-g_2(x,y)+h_1(x,y)+h_2(x,y)}}{-x
   ^2+x y+x-y}+\frac{2
   e^{-g_1(x,y)-g_2(x,y)+k_1(
   x,y)+k_2(x,y)}}{(y-1) (x-y)}\cr\cr&-\frac{2
   (x-1)
   e^{-h_1(x,y)-h_2(x,y)+k_1(
   x,y)+k_2(x,y)}}{(y-1)
   (x-y)^2}+\frac{2}{(x-y)^2}\cr
   \cr\cr
   \nonumber
   \end{align}
   \eject
   \begin{align}
\mathcal{D}_2 &= -\frac{8}{(x-y)^2}+ \frac{e^{-2 (g_1(x,y)+g_2(x,y))-2
   h_2(x,y)+2 k_2(x,y)} x^2
   \left(e^{2 g_1(x,y)}-e^{2
   g_2(x,y)}\right)^2}{X(x)
   Y(y)}\cr\cr&-\frac{3 e^{-2
   (g_1(x,y)+g_2(x,y))+h_1(x,
   y)-h_2(x,y)-k_1(x,y)+k_2(x
   ,y)} x y \left(e^{2 g_1(x,y)}-e^{2
   g_2(x,y)}\right)^2}{X(x)
   Y(y)}\cr\cr&-\frac{2
   e^{-g_1(x,y)-g_2(x,y)+h_1(
   x,y)+h_2(x,y)}}{(x-1) (x-y)}-\frac{11
   e^{-2 g_1(x,y)-2
   g_2(x,y)+h_1(x,y)+h_2(x,y)
   +k_1(x,y)+k_2(x,y)}}{(x-1)
   (y-1)}\cr\cr&+\frac{14
   e^{-h_1(x,y)-h_2(x,y)+k_1(
   x,y)+k_2(x,y)} (x-1)}{(x-y)^2
   (y-1)}+\frac{3
   e^{-g_1(x,y)-g_2(x,y)+k_1(
   x,y)+k_2(x,y)}}{y^2-x
   y-y+x}\cr\cr&+\frac{e^{-g_1(x,y)-g_2(x,
   y)-k_1(x,y)+k_2(x,y)}}{2 (y-1)^3
   X(x)}+\frac{e^{-2 g_1(x,y)-2
   g_2(x,y)+h_1(x,y)+h_2(x,y)
   -k_1(x,y)+k_2(x,y)} (x-y)}{2
   (x-1) (y-1)^3
   X(x)}\cr\cr&+\frac{e^{g_1(x,y)-3
   g_2(x,y)+h_1(x,y)-h_2(x,y)
   +2 k_2(x,y)} x (x-y) y}{(y-1)
   X(x) Y(y)}\cr\cr&-\frac{2
   e^{-g_1(x,y)-g_2(x,y)+h_1(
   x,y)-h_2(x,y)+2 k_2(x,y)} x
   (x-y) y}{(y-1) X(x)
   Y(y)}\cr\cr&
   +\frac{e^{-3
   g_1(x,y)+g_2(x,y)+h_1(x,y)
   -h_2(x,y)+2 k_2(x,y)} x (x-y)
   y}{(y-1) X(x)
   Y(y)} +\frac{6
   e^{-2 g_1(x,y)-2 g_2(x,y)+2
   k_1(x,y)+2
   k_2(x,y)}}{(y-1)^2}\cr\cr&
\end{align}
\eject
\begin{align}
&   -\frac{4 e^{-3
   g_1(x,y)-3
   g_2(x,y)+h_1(x,y)+h_2(x,y)
   +2 k_1(x,y)+2 k_2(x,y)}
   (x-y)}{(x-1) (y-1)^2}\cr\cr&+\frac{4
   e^{-g_1(x,y)-g_2(x,y)-h_1(
   x,y)-h_2(x,y)+2 k_1(x,y)+2
   k_2(x,y)} (x-1)}{(x-y)
   (y-1)^2}\cr\cr&-\frac{6 e^{-2 h_1(x,y)-2
   h_2(x,y)+2 k_1(x,y)+2
   k_2(x,y)} (x-1)^2}{(x-y)^2
   (y-1)^2} +\frac{3 e^{3 g_1(x,y)-3
   g_2(x,y)+h_1(x,y)-3
   h_2(x,y)+2 k_2(x,y)} x (x-y)
   y^2}{X(x) Y(y)^2}\cr\cr&+\frac{3
   e^{-3 g_1(x,y)+3
   g_2(x,y)+h_1(x,y)-3
   h_2(x,y)+2 k_2(x,y)} x (x-y)
   y^2}{X(x) Y(y)^2}\cr\cr
&+\frac{3
   e^{g_1(x,y)-g_2(x,y)+h_1(x
   ,y)-3 h_2(x,y)+2 k_2(x,y)} x y^2
   (y-x)}{X(x) Y(y)^2}+\frac{3
   e^{-g_1(x,y)+g_2(x,y)+h_1(
   x,y)-3 h_2(x,y)+2 k_2(x,y)} x
   y^2 (y-x)}{X(x) Y(y)^2}\cr \nonumber
\end{align}

\section{Equations for  NS5 on a cone over $Y^{p q}$}\label{app:cone}

In this appendix we present the complete  set of equations for  $NS5$ branes wrapping a two cycle on the cone over $Y^{p,q}$. From the calibrating conditions we obtain:

\begin{eqnarray}
\partial_r \phi(r,y)&=& \frac{e^{k_2(r,y)-k_1(r,y)} \left(e^{-{g_1}(r,y)-{g_2}(r,y)}-e^{-{h_1}(r,y)-{h_2}(r,y)}\right)}{2 (c y-1)^2}  \cr\cr
\partial_r g_1(r,y)&=&\frac{e^{-g_1(r,y)-g_2(r,y)-{k1}(r
   ,y)+k_2(r,y)}}{2 (c y-1)^2}  \cr\cr
\partial_r g_2(r,y)&=& \frac{e^{-g_1(r,y)-g_2(r,y)-k_1(r
   ,y)+k_2(r,y)}}{2 (c y-1)^2} \cr\cr
\partial_r h_1(r,y)&=& \frac{\left((c y-1)^2 e^{2 k_1(r,y)}-1\right)
   e^{-h_1(r,y)-h_2(r,y)-k_1(r,y)
   +k_2(r,y)}}{2 (c y-1)^2} \cr\cr
\partial_r h_2(r,y)&=& \frac{\left((c y-1)^2 e^{2 k_1(r,y)}-1\right)
   e^{-h_1(r,y)-h_2(r,y)-k_1(r,y)
   +k_2(r,y)}}{2 (c y-1)^2} \cr\cr
\partial_r k_1(r,y)&=& \frac{e^{-g_1(r,y)-g_2(r,y)-k_1(r
   ,y)+k_2(r,y)}}{2 (c
   y-1)^2} -\frac{e^{-h_1(r,y)-h_2(r,y)-
   k_1(r,y)+k_2(r,y)}}{2 (c
   y-1)^2}\cr &&+\frac{1}{2}
   e^{g_1(r,y)-g_2(r,y)-k_1(r,y)+
   k_2(r,y)}+\frac{1}{2}
   e^{-g_1(r,y)+g_2(r,y)-k_1(r,y)
   +k_2(r,y)}\cr &&-\frac{1}{2}
   e^{-g_1(r,y)-g_2(r,y)+k_1(r,y)
   +k_2(r,y)}-\frac{1}{2}
   e^{-h_1(r,y)-h_2(r,y)+k_1(r,y)
   +k_2(r,y)}\cr
\cr
\partial_y \phi(r,y)&=&0 \qquad   \partial_y k_1(r,y)= 0  \qquad
\partial_y k_2(r,y)= 0  \cr  \cr
\partial_y g_1(r,y)&=& -\frac{3 y (c y-1)
   e^{g_1(r,y)-g_2(r,y)+h_1(r,y)-
   h_2(r,y)}}{2 \left(y^2 (2 c
   y-3)+w\right)}+\frac{3 y (c y-1)
   e^{-g_1(r,y)+g_2(r,y)+h_1(r,y)
   -h_2(r,y)}}{2 \left(y^2 (2 c
   y-3)+w\right)}\cr &&+\frac{c
   e^{-g_1(r,y)-g_2(r,y)+h_1(r,y)
   +h_2(r,y)}}{2 c y-2}+\frac{c}{2-2 c y} \cr\cr
\partial_y g_2(r,y)&=& \frac{3 y (c y-1)
   e^{g_1(r,y)-g_2(r,y)+h_1(r,y)-
   h_2(r,y)}}{2 \left(y^2 (2 c
   y-3)+w\right)}-\frac{3 y (c y-1)
   e^{-g_1(r,y)+g_2(r,y)+h_1(r,y)
   -h_2(r,y)}}{2 \left(y^2 (2 c
   y-3)+w\right)}\cr &&+\frac{c
   e^{-g_1(r,y)-g_2(r,y)+h_1(r,y)
   +h_2(r,y)}}{2 c y-2}+\frac{c}{2-2 c y} \cr\cr
\partial_y h_2(r,y)&=& \frac{3 y (c y-1)
   e^{g_1(r,y)-g_2(r,y)+h_1(r,y)-
   h_2(r,y)}}{2 \left(y^2 (2 c
   y-3)+w\right)}+\frac{3 y (c y-1)
   e^{-g_1(r,y)+g_2(r,y)+h_1(r,y)
   -h_2(r,y)}}{2 \left(y^2 (2 c
   y-3)+w\right)}\cr &&+\frac{c
   e^{-g_1(r,y)-g_2(r,y)+h_1(r,y)
   +h_2(r,y)}}{2-2 c y}+\frac{-4 c^2 y^3+c
   \left(w+9 y^2\right)-6 y}{2 (c y-1) \left(y^2 (2
   c y-3)+w\right)}                           \cr\cr
\cr \nonumber
\end{eqnarray}

Since we are dealing with PDE's we have to demand that $\partial_r \partial _y = \partial _y \partial_r$.  From this integrability requirement we obtain two more equations and two algebraic constraints,

\begin{eqnarray}
\partial_r h_2(r,y) &=& \frac{\left((c y-1)^2 e^{2 k_1(r,y)}-1\right)
   e^{-h_1(r,y)-h_2(r,y)-k_1(r,y)
   +k_2(r,y)}}{2 (c y-1)^2}\cr\cr
\partial_y h_1(r,y)&=& \frac{c \exp (-2 g_1(r,y)-2 g_2(r,y)+2
   h_1(r,y)+2 h_2(r,y))}{c y-1}\cr &&-\frac{3
   y (c y-1)
   e^{g_1(r,y)-g_2(r,y)+h_1(r,y)-
   h_2(r,y)}}{2 \left(y^2 (2 c
   y-3)+w\right)}-\frac{3 y (c y-1)
   e^{-g_1(r,y)+g_2(r,y)+h_1(r,y)
   -h_2(r,y)}}{2 \left(y^2 (2 c
   y-3)+w\right)}\cr &&+\frac{3 c
   e^{-g_1(r,y)-g_2(r,y)+h_1(r,y)
   +h_2(r,y)}}{2 c y-2}+\frac{-4 c^2 y^3-5 c
   w+3 c y^2+6 y}{2 (c y-1) \left(y^2 (2 c
   y-3)+w\right)},   \nonumber
\end{eqnarray}

\begin{eqnarray}
\mathcal{C}_1 &=& -2 c \left(y^2 (2 c y-3)+w\right) \left((c y-1)^2
   e^{2 k_1(r,y)}-4\right) e^{g_1(r,y)+g_2(r,y)+2 h_1(r,y)+2 h_2(r,y)+k_2(r,y)}\cr &&-4 c \left(y^2 (2 c y-3)+w\right) \left((c y-1)^2 e^{2 k_1(r,y)}-1\right)e^{(2 g_1(r,y)+2 g_2(r,y)+h_1(r,y)+h_2(r,y)+k_2(r,y))}\cr &&-3 y (c y-1)^4 e^{5 g_1(r,y)+g_2(r,y)+2 h_1(r,y)+k_2(r,y)}+6 y (c y-1)^4 e^{3 g_1(r,y)+3 g_2(r,y)+2 h_1(r,y)+k_2(r,y)}\cr &&-3 y (c y-1)^4 e^{g_1(r,y)+5 g_2(r,y)+2 h_1(r,y)+k_2(r,y)}\cr &&+2 c (c y-1)^2 \left(y^2 (2 c y-3)+w\right) e^{3 g_1(r,y)+3 g_2(r,y)+2 k_1(r,y)+k_2(r,y)}\cr &&+4 c \left(y^2 (2 c y-3)+w\right) \left((c y-1)^2 e^{2 k_1(r,y)}+1\right) e^{3 h_1(r,y)+3 h_2(r,y)+k_2(r,y)} \cr \cr
\mathcal{C}_2 &=& -6 y (c y-1)^2 e^{2 g_1(r,y)+2 g_2(r,y)+2 h_1(r,y)+k_2(r,y)}\cr &&+2 c \left(y^2 (2 c y-3)+w\right) e^{2 g_1(r,y)+2 g_2(r,y)+2 k_1(r,y)+k_2(r,y)}\cr &&-2 c \left(y^2 (2 c y-3)+w\right) e^{2 h_1(r,y)+2
   h_2(r,y)+2 k_1(r,y)+k_2(r,y)}+3 y (c y-1)^2 e^{4 g_1(r,y)+2 h_1(r,y)+k_2(r,y)}\cr &&+3 y (c y-1)^2 e^{4 g_2(r,y)+2 h_1(r,y)+k_2(r,y)}\cr \nonumber
\end{eqnarray}

\newpage
\bibliographystyle{JHEP}
\bibliography{Ypqbib}

\end{document}